\documentclass[aps,prc,onecolumn,showpacs,superscriptaddress,groupedaddress]{revtex4}

\usepackage{amsmath}
\usepackage{amssymb}
\usepackage{amsfonts}
\usepackage{float}
\usepackage{graphicx}
\usepackage{tabularx}
\usepackage{multirow}
\usepackage{lscape}
\usepackage{rotating}
\usepackage{pstricks}
\usepackage{enumitem}

\begin{document}

\title{A new statistical method for the structure of the inner crust of neutron stars}


\author{A. Pastore}
\affiliation{Department of Physics, University of York, Heslington, York, Y010 5DD, UK}

\author{M. Shelley}
\affiliation{Department of Physics, University of York, Heslington, York, Y010 5DD, UK}

\author{S. Baroni}
\affiliation{Independent researcher, Barcelona, ES}    

\author{C. A. Diget}
\affiliation{Department of Physics, University of York, Heslington, York, Y010 5DD, UK}         

\begin{abstract}

We investigated the structure of the low density regions of the inner crust of neutron stars using the Hartree-Fock-Bogoliubov (HFB) model
to predict the proton content $Z$ of the nuclear clusters
and, together with the lattice spacing, the proton content of the crust as a function of the total 
baryonic density $\rho_b$.
The exploration of the energy surface in the $(Z,\rho_b)$ configuration space
and the search for the local minima require thousands of calculations. Each of them implies
an HFB calculation in a box with a large number of particles, thus making the whole process very demanding.
In this work, we apply a statistical model based on a Gaussian Process Emulator 
that makes the exploration of the energy surface 
ten times faster. We also present a novel treatment of the HFB equations that leads to an uncertainty on the total energy 
of $\approx 4$ keV per particle.
Such a high precision is necessary to distinguish neighbour configurations around the energy minima.

\end{abstract}


\pacs{ XXX}
 
\date{\today}


\maketitle

%
\section{Introduction}
\label{sect:intro}
Pulsars are celestial objects that emit a highly periodic radiation beam in the radio and x-ray wavelengths. 
Since the first pulsar was discovered in 1967~\cite{hew68},
more than 1500~\cite{hul94,man01} have then been observed and it has been estimated that there are probably 25000 potentially 
observable pulsars in the Galaxy beamed toward the Earth ~\cite{manchester2004pulsars}.
They are believed to be neutron stars rotating at the observed pulse period, 
which ranges from $\approx$11 s for the normal pulsars down to $\approx$1.4 ms for the so-called millisecond pulsars~\cite{lattimer2004neutronstars}.

In effect, neutron stars are the only possible candidates for pulsars~\cite{mic91}, 
as such short spin periods are attainable only by extremely dense rotating stars,
where the strong gravitational field prevents the star from ejecting mass into the surrounding space.
The typical mass and radius of neutron stars are $M = 1.5 M_{\odot}$ and $R=12$ km ~\cite{hae07} (where  $M_{\odot}$ is the solar mass), 
which gives an average density of the order of $10^{14}$ g/cm$^{3}$.
It is the same order of magnitude of the nuclear saturation density $\rho_0=0.16$ fm$^{-3}=1.6\cdot 10^{38}$cm$^{-3}\approx 2.7\cdot 10^{14}$g/cm$^{3}$.
Now, a simple order-of-magnitude estimate shows that this density is large enough to sustain millisecond spin periods
\cite{reisenegger2016ordermagnitude},
as the average density $\bar{\rho}$ of a star and the minimum spin period achievable by the star relate as follows:
\begin{eqnarray}\label{eq:order-of-magn}
  P_{min}\approx \left(\frac{3\pi}{G\bar{\rho}}\right)^{1/2}\;,
\end{eqnarray}
\noindent where $G$ is the gravitational constant. For the observed period $P \approx 1.4$ ms of millisecond pulsars, this gives an estimate for the density of the pulsar of $\bar{\rho}\approx 10^{14}$g/cm$^3$, 
in agreement with the average density of a neutron star.

Now, such a large average density implies a central baryonic density in the core of neutron stars ranging between $3\rho_0$ and $10\rho_0$, which leads to the conclusion that neutron stars are the densest cold baryonic matter outside of black holes
~\cite{lattimer2005ultimate}. 
To see this, consider the following:
\begin{enumerate}[label=(\alph*)]

\item equation-of-state (EOS)-independent analytic solutions of Einstein's 
equations can be used to set an upper limit on the maximum density 
for cold baryonic matter, based on the largest measured neutron star mass~\cite{lattimer2005ultimate}

\item the recently discovered J1164-2230 millisecond pulsar \cite{demorest2010twosolarmasses} with a mass $M=(1.97 \pm 0.04)M_\odot$ and the J0348+0432~\cite{antoniadis2013massive} with a mass $M=(2.01 \pm 0.04)M_\odot$,
the heaviest ever discovered \cite{fonseca2016nanograv},
set this upper limit to $(3.74\pm 0.15)\cdot 10^{15}$g/cm${^3}\approx 10\rho_0$. 

\end{enumerate}
\noindent The quest of the EOS to describe the properties of these macroscopic objects represents one of the major challenges in nuclear physics.
Theoretical models predict different EOSs for baryonic matter at such extreme densities. 
Some of these models include the presence of deconfined quarks or hadronic exotic states, 
as the formation of kaons or hyperons ~\cite{lattimer2007neutronstars,glendenning1998kaons,lackey2006hyperons},
but almost all of the models have difficulties in explaining the heavy mass of J1164-2230~\cite{demorest2010twosolarmasses}.  
The main reason is because these exotic degrees of freedom lower the Fermi energy of the baryonic matter 
in the core, depressing the baryonic pressure and hence the maximum allowed mass of the star~\cite{vid00}.
Although recent groups have found some possible solution to solve this problem~\cite{ast14,lon15}, the properties of the inner core~\cite{cha08} of the star still represent a major challenge for current theoretical models. For a more detailed discussion on the subject we refer to the recent review article ~\cite{cha16}.

While the composition of the core of neutron stars is still unclear, there is a better agreement on the structure of the outer layers~\cite{cha08}. 
Right outside of the core, in the so-called \emph{crust}, the baryonic density already drops to infra-nuclear values 
$\rho_b \approx \frac{\rho_0}{2}$~\cite{douchin2000inner} 
and gradually decreases to $\approx 10^{-8}\rho_0$ in the outer regions of the crust. 
The outmost layers, the \emph{envelope} and the \emph{atmosphere}, contain a negligible amount of mass, 
although they play a fundamental role in the energy transport and photoemission of the surface of the 
star~\cite{lattimer2004neutronstars}. 

The crust extends for 1 to 2 km and it is made of a periodic lattice of nuclei, 
whose proton content and relative distance vary with the total baryonic density.
These nuclei are surrounded by a gas of ultra-relativistic electrons. In addition,
in the crust regions closer to the core, the density is larger than the neutron drip 
density $4.7\cdot 10^{11}$g/cm$^3 \approx 10^{-3}\rho_0$
and neutrons leak out of the nuclei. This is the so-called \emph{inner crust} of neutron stars: 
it is made of nuclei immersed in a superfluid neutron gas, together with an ultra-relativistic electron gas.

Although neutrons in vacuum undergo $\beta$ decay with a half life of 15 minutes, 
in dense matter they are stabilised, thanks to the Pauli exclusion principle. 
To understand why, one must consider that charge neutrality requires that
the electron density $\rho_e$ equals the proton density $\rho_p$. 
At these high densities, the electrons are highly relativistic. 
This means that even if the inner crust is at zero temperature~\cite{cha08}, the electrons are energetic enough for the inverse
decay to take place, $i.e.$ a proton capturing an electron and producing a neutron.
If the neutron chemical potential is larger than the proton plus electron chemical potentials,
$\mu_n > \mu_p + \mu_e$,
 the neutrons will decay more frequently than they are created in the inverse $\beta$ decay. 
The opposite happens if
$\mu_n < \mu_p + \mu_e$, until the following equilibrium is reached
\begin{equation}\label{eq:beta_equilibrium_equal}
  \mu_n = \mu_p + \mu_e\;.
\end{equation}

\noindent Assuming that protons and neutrons form a gas in  $\beta$-equilibrium with the electrons, Eq.\ref{eq:beta_equilibrium_equal} can be expressed as

\begin{eqnarray}
\mu_n-\mu_p=-\left.\frac{\partial E/A}{\partial x}\right|_{\rho_b}\;,
\end{eqnarray}

\noindent where $x=\rho_p/\rho_b=Z/A$ is the proton content of the system and the derivative of the equation of state (E/A) is taken at constant baryonic density. Using a parabolic approximation then we have~\cite{baldo1997microscopic}

\begin{eqnarray}
\mu_n-\mu_p=4 E_{sym}(\rho_b)(1-2x)\;,
\end{eqnarray}

\noindent showing that the proton/neutron asymmetry strongly depends on the symmetry energy $E_{sym}(\rho_b)$ of the system. The latter can be extracted from nuclear physics models~\cite{chen2005determination,centelles2009nuclear,gandolfi2012maximum,hebeler2014symmetry}.

The typical values of $x$ vary between $x\approx10^{-2}$ for $\rho_b\approx\rho_0/3$ to $x\approx0.15$ in the neutron star interior. See Ref.~\cite{baldo1997microscopic} for more details.
This estimate is not valid in the region of the inner crust of the star, since in these region neutrons and protons do not exist as free particles, but they do form clusters~\cite{ducoin2008cluster}. We need to treat inhomogeneous matter to properly determine the proton fraction.

The aim of this work is thus to determine the proton content of the inner crust using more advanced nuclear structure models.
The inner crust  plays a crucial role to understand several phenomena related to the physics of the whole neutron star,
as for example the cooling process~\cite{pethick1995matter,yakovlev2004neutron}. 
Given the very peculiar structure of the inner crust, the tool of choice to obtain an accurate description of this region of the star is based on the nuclear energy density functional (NEDF) theory~\cite{bender2003self}. This method is well suited to describe properties of nuclei along the nuclear chart from drip-line to drip-line and from light to heavy elements with remarkable accuracy~\cite{gor13b}. 
The first application of NEDF to the inner crust dates back to the pioneering work of Negele and Vautherin~\cite{Negele1973}. By using the Wigner-Seitz (WS) approximation~\cite{wig33}, they  have been able to determine the cluster composition of the crust at different values of the baryonic density.
In this work, the authors did not consider pairing correlations~\cite{brink2005nuclear}. Later works~\cite{bal05,gri11} have tried to quantify the effect of pairing correlations on the cluster structure of the star, by solving the fully self-consistent Hartree-Fock-Bogoliubov (HFB) equations for some specific functional. Unfortunately the results seemed to suffer from numerical noise, mainly due to the treatment of the neutron gas states~\cite{bal06}.
The treatment of the gas states has thus become a major issue in these type of calculations and most of the groups have decided to overcome such a problem by using semi-classical methods~\cite{brack1985selfconsistent,pea15}. 
Although these methods rely on several approximations, for example they do not consider pairing correlations, they have the advantage to avoid artificial shell effects in the neutron  gas and they are computationally less demanding than a full HFB calculation.

In the present article, we present a systematic investigation of the possible sources of numerical noise for HFB calculations, with a special attention to the shell effects of the neutron gas states, together with a new statistical method to reduce the computational effort required for these calculations.
The article is organised as follows: section \ref{the_method} explains how the HFB model can be used to search for the minimum energy configurations
of the inner crust. In section \ref{GPE} we outline the features of the new statistical method that we used to explore the energy
surface, while in section \ref{larger_box} we discuss a novel treatment of the states in the continuum,
which allows us to greatly suppress the error on the HFB total energy.
The different contributions to the HFB total energy are discussed in Sections \ref{total_energy},
\ref{nuclear_energy}, \ref{electron_energy}, \ref{ep_energy} and \ref{sect:beta_eq}, together with a discussion
of the sources of error arising from the approximations and limitations of the model.
The results of our calculations for the inner crust are presented in Section \ref{cluster_structure}.
We finally draw our conclusions in Sec.~\ref{sec:concl}.

\section{The method}\label{the_method}

We adopt the so-called Wigner-Seitz  cell approximation \cite{wig33}, 
which divides the periodic lattice of nuclei of the inner crust of neutron stars
into spherical charge-neutral non-interacting cells.
The entire structure of the crust is obtained by iterating a spherical cell to all directions.
Periodic boundary conditions at the border of the box ensure continuity when
passing from a cell to its neighbours.
For the validity of the WS approximation, we refer to Ref.~\cite{cham07}.
For a given total baryonic density $\rho_b$, the cells are all identical and each one contains
one nuclear cluster at its center. The radius $R_{WS}$ of a cell is then half of the distance between
neighbouring clusters.
Each cell also contains an ultra-relativistic electron gas that can be considered uniform to a good approximation
(see Section \ref{electron_energy}) plus a gas of superfluid neutrons surrounding the cluster. 

As charge neutrality requires that the number of electrons equals the number of protons,
a WS cell can then be defined by its radius $R_{WS}$ and the neutron and proton numbers $N$ and $Z$.
Alternatively, if the total baryonic density $\rho_b$ and the size of the cell are given,
the total number of nucleons $A=N+Z=\rho_b \cdot \frac{4}{3}\pi R_{WS}^3$ is also given.
A cell can thus be uniquely defined by $(Z, \rho_b, R_{WS})$. The proton content of a cell, and hence
the proton content of the density region that the cell represents, is given by
\begin{equation}\label{eq:proton content of a cell}
  Y(Z, \rho_b, R_{WS}) = \frac{Z}{A} = \frac{Z}{\rho_b \cdot \frac{4}{3}\pi R_{WS}^3}\;.
\end{equation}

\noindent In principle, the whole $(Z, \rho_b, R_{WS})$ configuration space should be explored in search for
the energetically favoured configurations, corresponding to the minima of the energy surface 
$E_{tot}(Z, \rho_b, R_{WS})$. $E_{tot}$ is the total energy of a WS cell, as discussed in details in
Section \ref{total_energy}.
The set of these minima describes a path in the $(Z, \rho_b, R_{WS})$ configuration space,
hence defining in details the evolution of the structure of the inner crust of neutron stars from its outmost
to its inmost regions.
In the current article, we limit ourselves to the zero-temperature case, but the formalism can be extended to include 
the non-zero temperature case~\cite{onsi2008semi,pas15a}.

We investigate a region of the baryonic density that spans $0.0004$ fm$^{-3}\leq \rho_b \leq0.02$ fm$^{-3}$. This region is actually smaller than the actual inner crust that goes from  $10^{-3} \rho_0 \approx 0.00016$ fm$^{-3}$ to
$\frac{1}{2}\rho_0\approx0.08$ fm$^{-3}$~\cite{pea12}.
We excluded the drip-line region, since a recent work~\cite{chamel2015neutron} shows that the neutron-drip
transition could occur at lower densities than what mean-nucleus approaches could predict. The high density region has also been excluded since the numerical noise of our calculations is too large and the results would not be reliable.

%

For what concerns the proton number, we consider the range $16\leq Z \leq 60$, as clusters
 outside of this range have been shown to be much more unstable~\cite{bal05,gri11,pea12,grill2012neutron,grill2014equation}.

When performing the minimisation of $E_{tot}(Z, \rho_b, R_{WS})$ using HFB equations, we have to deal with two main limitations.
\begin{enumerate}[label=(\alph*)]
\item To calculate the properties of a WS cell, we need to impose boundary conditions that could lead to an artificial discretisation of the states in the
continuum, which translates into a systematic uncertainty on the total energy of the cell.
As discussed in Sect.~\ref{larger_box}, 
this non-physical source of error can be as large as $\approx 200$ keV per particle~\cite{mar07}.
\item The small energy difference between configurations around the minima requires a fine
scan of the $(Z, \rho_b, R_{WS})$ configuration space.
This, though, implies a very large number of WS cells to be calculated, and as each of them
is already computationally demanding, the whole exploration of the
energy surface becomes prohibitive.
\end{enumerate}
In this work we present a simple solution for both problems.
\begin{enumerate}[label=(\alph*)]
\item First, in order to reduce the uncertainty on the HFB total energy, the calculations
are carried out in a box of radius $R_B > R_{WS}$.
As discussed in Section \ref{larger_box}, 
this suppresses the effect of the non-physical discretisation of the states in the continuum,
hence reducing the associated uncertainty on the HFB total energy to $\approx 1$ keV per particle.
Although more advanced HFB methods are now available to properly treat continuum states ~\cite{chamel2005band,gogelein2007nuclear,chamel2012neutron,sch15,shi17}, it is still unclear if these methods can deal with very large WS cells especially in terms of execution time. We leave this analysis for a forthcoming project.
Using a radius of the box $R_B$ larger than the WS radius $R_{WS}$ and
treating the  proton-electron interaction in the WS cell perturbatively, we downsize the configuration space
from three to two dimensions, as the WS radius $R_{WS}$ is no longer an independent
variable and a configuration is uniquely defined by $(Z,\rho_b)$ only.
The details of this method are discussed in Sect.~\ref{larger_box}.

Once the HFB neutron and proton density profiles $\rho_{q=n,p}$ are computed for a given
cell $(Z,\rho_b)$, the WS cell radius $R_{WS}$ is found by imposing $\beta$-equilibrium and the proton and neutron numbers are obtained as 

\begin{equation}\label{calculate_RWS}
  \int_0^{R_{WS}}dr 4\pi r^2\rho_q(r)=N_q\;.
\end{equation}

\item Although the above approximation decreases dramatically the number of configurations to be surveyed,
for the exploration of the surface to be feasible, the adoption of a new computational method 
that further reduces the number of configurations to be computed is necessary.

As a matter of fact, a statistical model called Gaussian Process Emulator (GPE) can be used 
to reduce the number of configurations by one order of magnitude.
The features of the GPE and how it can boost the exploration of the energy surface
are outlined in Section \ref{GPE}.
\end{enumerate}


\subsection{States in the continuum}\label{larger_box}

To solve the HFB problem in a WS cell, one need to impose boundary conditions (BC). A common choice is to use the Dirichlet-Neumann mixed boundary conditions~\cite{Negele1973}. This leads to a discretisation of the states in the continuum, with an artificial spacing between the eigenstates of the neutron gas which is proportional to $\approx\frac{1}{R_{WS}^2}$~\cite{messiah1958quantum}. This does not represent the reality of the crust~\cite{cham07},
it affects the binding energy of the system and it is also responsible for a non-physical dependence
of the proton content of the clusters on the size $R_{WS}$ of the cell~\cite{bal06}.
While this effect is negligible at low densities, 
it becomes more and more important in the high density regions of the inner crust,
where the clusters get closer to each other.

As suggested in Ref.~\cite{mar07}, a simple way to estimate this effect is to 
check the Hartree-Fock (HF) results for pure neutron matter (PNM) in a box, against an analytical solution.
We thus calculated 
the energy per particle $\left.\frac{E}{N}\right|_{HF}$ 
of a piece of uniform pure neutron matter as a function of the box radius $R_B$, 
and compared it with the exact analytical solution $\left.\frac{E}{N}\right|_{exact}$ ~\cite{davesne2016extended}.
It is worth recalling that there are two possible definitions of Dirichlet-Neumann BC. We have checked that the results do not depend on which one is used. The first is called boundary conditions even (BCE): 
(i) even-parity wave functions vanish at $R_B$; (ii) the first derivative of odd-parity wave functions vanishes at $R_B$.
The other is called boundary conditions odd (BCO), where the two parity states are treated in the opposite way. We refer to Ref.~\cite{pas11} for more details.
The error induced on the energy per particle by the state discretisation can then be defined as

\begin{equation}\label{err:HF}
  \delta e(R_B)= \left.\frac{E}{N}\right|_{exact}-\left.\frac{E}{N}\right|_{HF}(R_B)\;.
\end{equation}

\begin{figure*}[!h]
\begin{center}
\includegraphics[width=0.40\textwidth,angle=-90]{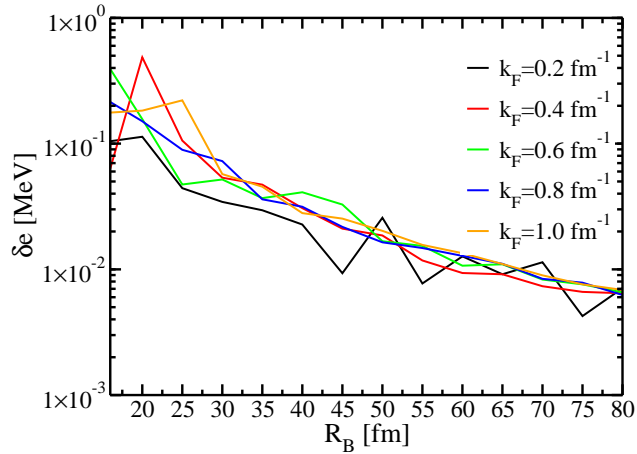}
\end{center}
\caption{(Colors online) The error on the energy per particle for PNM
as the difference between the HF energy calculated in a box of radius
$R_B$ and the analytical solution for PNM. See Eq.\ref{err:HF}.}
\label{HF_PNM}
\end{figure*}

\noindent The error $\delta e(R_B)$ is shown in Fig.\ref{HF_PNM} as a function of the box radius $R_B$. 
 As expected, the error decreases with the size of the box.
It decreases as  $\approx\frac{1}{R_{B}^2}$ and it is of the order of hundreds of keVs for small boxes,
while it reduces to $\approx 7$ keV for boxes with a radius $R_B=80$ fm.

As Fig.~\ref{HF_PNM} shows, the error $\delta e(R_B)$ also presents a dependence on the neutron density $\rho_n$
or, equivalently, on the neutron Fermi momentum $k_F=(3\pi^2 \rho_n)^{1/3}$, which
is not constant over the configuration space and contributes in 
different amounts to different density regions. It is then a source of error for the HFB total energy.
This error is less pronounced for larger boxes and it is $\approx 1$ keV per particle
for a box of 80 fm: this estimate comes from the spread of the curves in Fig.\ref{HF_PNM} at  $R_B=80$ fm.  In the rest of the article, we will make use of a constant box size of $R_B=80$ fm; as consequence the constant shift of $\approx 7$ keV per particle observed in Fig.\ref{HF_PNM}  will be roughly constant for
all the $(Z,\rho_b)$ configurations. Since we are interested in finding the minima of the $E_{tot}(Z,\rho_b)$ 
energy surface, a constant shift would not affect the calculations and it could be ignored. 

\begin{figure*}[!h]
\begin{center}
\includegraphics[width=0.40\textwidth,angle=-90]{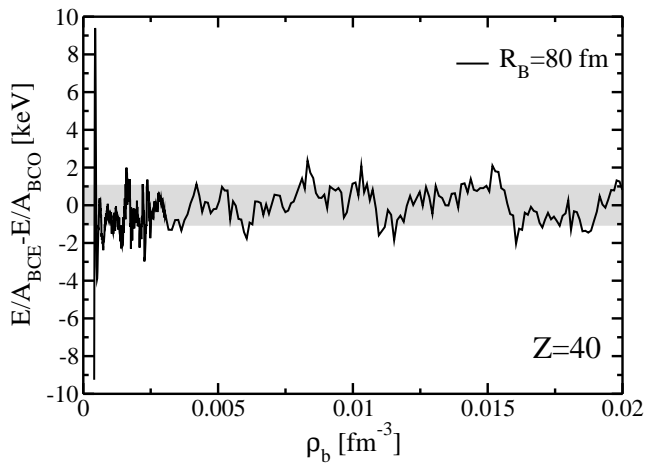}
\end{center}
\caption{(Colors online) Difference of energy per particle calculated at different baryonic densities for the cluster Z=40 and using two sets of boundary conditions (namely, BCE and BCO, see text for details). The grey band stands for $\pm1$ keV error band.}
\label{HF_Z40_oddeven}
\end{figure*}

Although the above estimate has been found for homogenous matter, we found a similar value for
inhomogeneous matter. We performed several calculations of the cluster with $Z=40$ at different values of baryonic density and for a box radius of $R_B=80$ fm. Notice that for these calculations we do not impose $\beta$-equilibrium since we are interested only on the nuclear contribution to the total energy. Different boundary conditions (i.e., BCE or BCO) give slightly different gas densities and hence different energies for the system. These energy differences
are then an estimate of the dependence of the error $\delta e(R_B)$ on the gas density. As it can be seen
in Fig.\ref{HF_Z40_oddeven}, in most of the cases the energy differences for $R_{B}=80$ are within 1 keV, thus confirming our analysis in the infinite medium. We expect similar results for other cluster compositions $Z$.
We notice that the use use of large boxes induces new numerical noise at very low-density. This error is still small compared to the typical error scale on the crust at this value of $\rho_b$, but it means that a further increase of the box would not be beneficial for the calculations.

We call $\sigma_{box}\approx 1$ keV per particle the error arising from the discretisation of the states in the
continuum.
It is important to bear in mind that this value should be considered as an upper-value since pairing correlations could help smearing out spurious shell effects, thus further reducing this error. Anyhow, we do not expect pairing correlations to reduce this error of one order of magnitude. To confirm such a statement, more detailed investigations are necessary.

In conclusion, these results suggest that the usage of a large box can greatly suppress the error induced by the 
Dirichlet-Neumann BC, although this implies larger computational times, 
as the number of basis states included in the calculation increases with the size of the box.

Finally, we consider a more realistic case taken from the Negele and Vautherin series of WS cells. We computed the total energy for the
WS cell $^{1500}$Zr, whose parameters have been taken from Ref.~\cite{Negele1973},
for two different configurations. 
In the first, the size of the box is equal to the size of the WS cell, i.e. $R_{B}=R_{WS}=19.6$ fm.
In the second, we fix $R_{B}=80$ fm while leaving $R_{WS}=19.6$ fm, imposing that the total number of neutrons 
and protons $N_{q=n,p}$ is fixed within the size of the WS cell (see Eq.\ref{calculate_RWS}).
Since in this section we want to quantify the discretisation effects on the neutron gas state, we neglect for this calculation the presence of the electrons.
The total energy per particle was then calculated for both configurations as

\begin{equation}\label{eq:integra}
  \frac{E}{A}=\frac{1}{A}\int_0^{R_{WS}}\sum_{q=n,p}\mathcal{H}_q(r) d^3 r\;,
\end{equation}

\noindent where $\mathcal{H}_{q}(r)$ is the Skyrme functional~\cite{bender2003self}. 
The total energy for the case $R_B=R_{WS}$  is $E/A=5.784$ MeV, while for $R_B=80$ fm we got $E/A=5.891$ MeV,
which corresponds to a difference of roughly $~100$ keV per particle.
As for this system $k_{Fn}=1.08$ fm$^{-1}$~\cite{pas11}, the result is compatible with the estimate done in Fig.\ref{HF_PNM}.

\begin{figure}[!h]
\begin{center}
\includegraphics[width=0.40\textwidth,angle=-90]{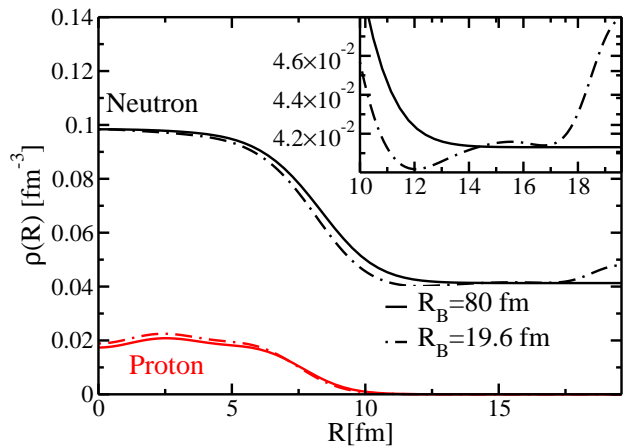}
\end{center}
\caption{(Colors online). Density profiles for the WS cell $^{1500}$Zr and two different size of the box. 
The inset is a zoom of the region of the cell outside the nuclear cluster. See text for details.}
\label{zn1500}
\end{figure}

We also checked the effect of the state discretisation on the neutron and proton densities profiles. The result is presented in Fig.\ref{zn1500}. 

The density profile in the case of $R_{B}=19.6$ fm is quite irregular and has a visible edge effect.
This increase of the density at the edge is compensated by an artificial depletion around 10 fm. 
With a larger box, though, these irregularities disappear and the gas region has very small fluctuations of the order 
of $\delta \rho\approx2\times10^{-5}$ fm$^{-3}$, which represent a clear improvement.


In Ref. ~\cite{gri11} the authors suggested the use of a correction function to take into account the noise introduced by the discretisation. 
For our calculation at $R_B=R_{WS}=19.6$ fm, this function predicts a correction $\delta E$=192 keV, 
which is larger then the value we have obtained using a larger box. This means that the subtraction method introduced in Ref.~\cite{mar07} induces an extra random noise of the order of few tens of keV that becomes very difficult to control.


\subsection{Gaussian Process Emulator}\label{GPE}

The GPE is a statistical model of interpolation~\cite{bastos2009diagnostics} that can be used
for a function, in the current case the energy surface $E_{tot}(Z, \rho_b)$, whose
values are the output of a complex non-random calculation, here an HFB calculation.
Each single calculation normally has several input parameters and the output is expected
to vary smoothly with the input, although in an unknown way.
The GPE predicts the values of the energy surface at a given point, starting from 
computed HFB energy values in the neighbourhood of the point.
Hence, it is sufficient to compute the HFB energy of only a small number of points of the $(Z,\rho_b)$ surface and the GPE will interpolate between these points.
The reconstruction of the energy surface is of course subject to errors, as the GPE
has to guess the value of the interpolated points. On top of that, the computed HFB energy
values themselves have an uncertainty, coming from the approximations and limitations of the model.
Because of this, each value that the GPE predicts is associated with an uncertainty.

In the present article, we have decided to test the GPE against \emph{exact} results to check the reliability of the technique in this specific case. To this purpose we have created a very dense grid ($\approx 20000$ configurations) of the  $E_{tot}(Z, \rho_b)$ surface.
Each point corresponds to a given HFB calculation.
We use a small subset of these points $\approx8$\% to test the GPE procedure.
The algorithm works as follow
\begin{enumerate}[label=(\roman*)]
\item Random points in the configuration space are selected using
a Latin Hypercube Sampling (LHS) ~\cite{levy2010computer}. This is because the GPE
performances are much better if it starts from random points rather
than from a uniform grid.
\item The points are taken by the GPE adding an error bar according to the estimated HFB numerical noise.
\item The GPE uses these points to produce the whole surface energy with relative 90\% confidence bands~\cite{bastos2009diagnostics}
\item We compare the resulting GPE energies with the ones obtained using the whole HFB grid.
\end{enumerate}

\noindent Usually the above procedure should be iterative, as for the regions of the surface with the largest errors, additional points from the simulator (HFB) should be added and the process then repeated up to a desired accuracy.

\begin{figure*}[!h]
\begin{center}
\includegraphics[width=0.340\textwidth,angle=-90]{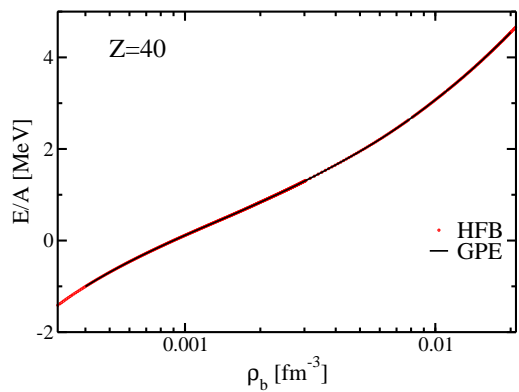}
\end{center}
\caption{(Colors online) Testing the accuracy of the GPE: total energy per particle for the
$(Z, \rho_b)$ configurations with $Z=40$. The line corresponds to the  GPE result using  a subset of data points ($\approx8$\%), while the dots correspond to the complete HFB data set. See text for details.}
\label{fig:benchmark_GPE_Z40}
\end{figure*}

The results are shown in Fig. \ref{fig:benchmark_GPE_Z40} for a one-dimensional cut for Z=40.
The confidence intervals of the GPE  are too small to be seen on this energy scale since they are of the order of $\pm5$ keV per particle.

\begin{figure}[!h]
\begin{center}
\includegraphics[width=0.50\textwidth,angle=0]{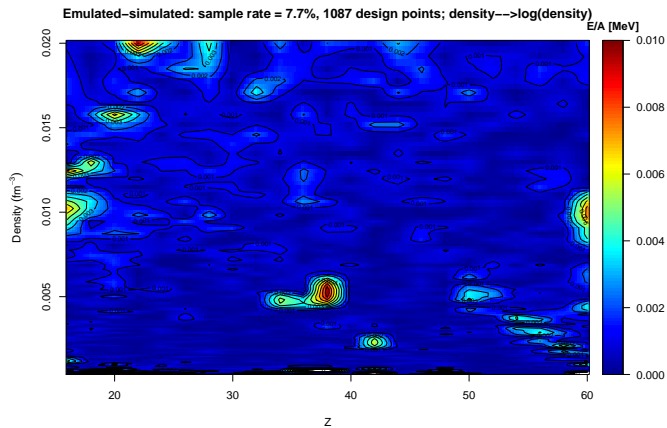}
\end{center}
\caption{(Colors online) Absolute energy difference between the computed HFB energies and the interpolated GPE values
for the whole $(Z, \rho_b)$ configuration space. See text for details.}
\label{fig:benchmark_GPE_all_space}
\end{figure}

To have a more global idea of the performances of the GPE on the whole $(Z,\rho_b)$ surface, we show in Fig.\ref{fig:benchmark_GPE_all_space} the absolute value of the difference between the interpolated surface and 
the actual HFB calculations.
On average, the difference between the \emph{exact} HFB results and the emulated ones is of the order of 2 keV per particle except for two regions: in the high density region for $Z\approx 20$ and 
in the low density regions with $Z\approx 36$. In these cases the deviation can grow up to 10 keV per particle.
This sudden increase is probably related to a sharp variation of the HFB output due to shell-effects. 
As discussed in the following section, these discrepancies are compatible with the current accuracy of the HFB code.

\section{HFB total energy of a WS cell}\label{total_energy}

The total energy of a WS cell with a given number of protons $Z$ and neutrons $N$ is given by~\cite{gri11}

\begin{eqnarray}\label{eq:etot}
E_{tot}=Z(m_pc^2+m_ec^2)+Nm_nc^2+E_{nuc}+E_{e}+E_{pe}\;,
\end{eqnarray}

\noindent where $m_{p,n,e}$ are the proton, neutron and electron masses. 
$E_{nuc}$ is the sum of the nuclear contribution and the Coulomb interaction between protons, 
$E_e$ is the energy contribution of the electron gas and $E_{pe}$ is the electromagnetic 
interaction between protons and electrons.
The minimisation at zero temperature of the total energy given in Eq.\ref{eq:etot} at fixed averaged baryonic density is equivalent to the minimisation of the Gibbs free energy per nucleon at a given pressure.
We refer the reader to  Ref.~\cite{onsi2008semi} for more details.
In Ref.~\cite{gulminelli2015unified}, the authors have discussed the possibility of first order phase transitions that break the single-nucleus approximation. We do not consider such a scenario in the present investigation, but this will be an important aspect to be considered when developing a unified equation of state which is able to give a realistic description of the crust of the NS.

Using the GPE method, $\rho_b$ can be safely considered as a continuous variable and we can easily compare different WS cells with the same baryonic density.

In the following, we give a detailed description of each term of Eq.\ref{eq:etot}.

\subsection{Nuclear energy $E_{nuc}$}\label{nuclear_energy}

To determine the nuclear energy contribution, we solve fully self-consistently the Hartree-Fock-Bogoliubov (HFB) equations, 
by projecting them on a spherical Bessel basis.
The equations read~\cite{Book:Ring1980}
\begin{eqnarray}\label{HFBeq}
  \sum_{n'}(h_{n'nlj}^q- \varepsilon_{F,q})U^{i,q}_{n'lj}+\sum_{n'}\Delta_{nn'lj}^qV^{i,q}_{n'lj}=E^{q}_{ilj}U^{i,q}_{nlj} & &
     ,\nonumber \\
  \sum_{n'}\Delta^q_{nn'lj}U^{i,q}_{n'lj} -\sum_{n'}(h^q_{n'nlj}- \varepsilon_{F,q})V^{i,q}_{n'lj}  =E^{q}_{ilj}V^{i,q}_{nlj} & &
     , 
\end{eqnarray}
where $\varepsilon_{F,q}$ is the Fermi energy and $q=n,p$ is the isospin index. 
We use the standard notation $nlj$ for the spherical single-particle states with radial quantum number $n$, 
orbital angular momentum $l$ and total angular momentum $j$.
$U^{i,q}_{nlj}$ and $V^{i,q}_{nlj}$ are the Bogoliubov amplitudes for the $i$-th quasiparticle of energy $E^{q}_{ilj}$.
In the limit of vanishing pairing gap $\Delta_{nn'lj}^q\rightarrow0$ one retrieves the usual Hartree-Fock equations on a basis.
More details on the solutions of these equations can be found in Refs~\cite{pas11,pas12,pastore2013pairing,pastore2015pairing}.

The matrix elements $h_{n'nlj}^q$ are calculated via the SLy4 Skyrme functional~\cite{cha97}, 
which includes the Coulomb interaction for protons ~\cite{ben05}.
For the pairing interaction, we adopt a density dependent delta interaction (DDDI) of the form~\cite{gar99}

\begin{eqnarray}\label{pair:int}
v^q_{pair}(\mathbf{r},\mathbf{r}')=V_0^q \left[ 1-\eta \left(\frac{\rho(\mathbf{R})}{\rho_0} \right)^\alpha\right] \delta(\mathbf{r}-\mathbf{r}')\;,
\end{eqnarray}

\noindent where $\alpha=0.45$, $\eta=0.7$ and $V_0=-430$ MeV$\cdot$fm$^{-3}$. 
These values have been tuned in Ref.~\cite{gar99} to reproduce the density dependence of the $^1$S$_0$~\cite{baroni2010partial} 
pairing gap in pure neutron matter with a realistic interaction at the BCS level.
To avoid the ultraviolet divergency~\cite{bul02}, we use a sharp cut-off of $E_{cut}=60$ MeV in the quasi-particle space. 

In the present article, we limit ourself to a simple mean-field calculation.  For such a reason we have decided to use an effective pairing interaction adjusted at BCS level in the infinite medium.
In principle, the effective interaction could be adjusted to incorporate effects beyond mean-field~\cite{gandolfi2008equation}. An example of such an adjustment has been done in Ref.~\cite{goriely2016further} to include self-energy effects into pairing interactions.


\subsection{Electron energy $E_e$}\label{electron_energy}

As the electrons are ultra-relativistic, one can consider them as a first approximation 
as decoupled from the protons and hence uniformly distributed in the WS cell.
Their energy contribution can be divided in kinetic $T_e$ and potential energy $U_e$.
The kinetic energy of an ultra-relativistic electron gas can be written as~\cite{sha93}

\begin{eqnarray}
T_e=Z m_e c^2\left\{ \frac{3}{8x^3}\left[x(1+2x^2) \sqrt{1+x^2}-\ln\left( x+\sqrt{1+x^2}\right)\right]-1\right\}\;;
\end{eqnarray}

\noindent where $x=\frac{\hbar k_{F_e}}{m_e c}$ and $k_{F_e}$ is the Fermi momentum of the electrons.  

Under the assumption of a uniform electron density, the electron-electron potential $V^{ee}_e$ is obtained 
as a sum of a direct and an exchange term. 
The latter has been calculated analytically for a relativistic Fermi gas in Ref.~\cite{sal61}. 
By integrating over the density, we thus get the potential energy contribution 

\begin{eqnarray}
U_e=\frac{3}{5} \frac{e^2Z^2}{R_{WS}}\left[ 1-\frac{5}{4} \left(\frac{3}{2\pi} \right)^{2/3} \frac{\Phi(x)}{Z^{2/3}} \right]\;,
\end{eqnarray}

\noindent where $\Phi(x) $ reads
\begin{eqnarray}
\Phi(x)=-\frac{1}{2x^2} \left[ 3x^2+x^4-6x \sqrt{1+x^2} \text{Sinh}^{-1}(x)+3\text{Sinh}^{-1}(x)^2\right]\;.
\end{eqnarray}

\noindent Since the screening length of the electrons is very large~\cite{maruyama2005nuclear}, the use of a constant electron density is fully justified. This approximation has been discussed in Refs~\cite{watanabe2003electron,alcain2014effect}.


\subsection{Proton-electron energy $E_{pe}$}\label{ep_energy}

Considering the Coulomb interaction between the protons in the central cluster of the WS cell and a uniform electron gas, we have the following
proton-electron potential term ~\cite{bon81}

\begin{eqnarray}\label{pot:ep}
V_{p-e}=-2\pi e^2 \rho_e \left( R_{WS}^2-\frac{1}{3}r^2\right) \;,
\end{eqnarray}

\noindent where $\rho_e$ is the electron density. 
The net effect of  the $V_{p-e}$ potential is then to reduce the diffusivity of the proton potential.

We tested the effect of this term by adding and removing it from the HFB calculation of the WS cell $^{1500}$Zr. This cell corresponds to an average baryonic density of $\rho_b=0.04$ fm$^{-3}$~\cite{pas12}.
We have decided to test our method in this cell since in this specific cell the potential given in Eq.\ref{pot:ep} has the larger effect. By inspecting more closely the equation and using $\rho_e=Z/\left( \frac{4}{3}\pi R_{WS}^3\right)$, we notice that $V_{p-e}$ increase linearly with the charge and with the inverse of the WS radius.
Notice that the cell $^{982}$Ge, which is the next cell in order of increasing density in the calculations of Negele and Vautherin, can not be considered since it is unstable~\cite{pas11}.

\begin{figure}[!h]
\begin{center}
\includegraphics[width=0.40\textwidth,angle=-90]{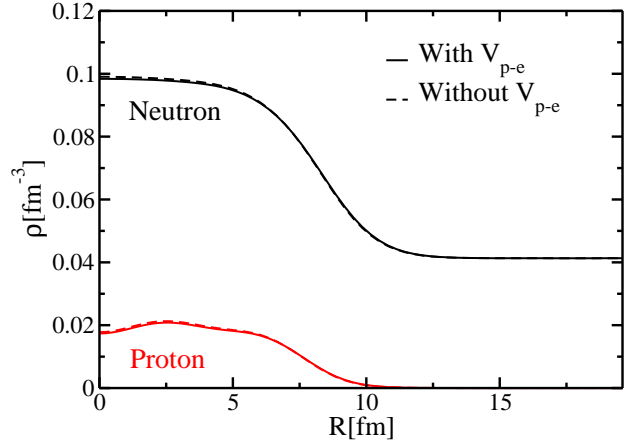}
\end{center}
\caption{(Colors online). Neutron and proton density profiles for the WS cell $^{1500}$Zr,
when the proton-electron $V_{p-e}$ potential is included or neglected in the 
total energy of the system. See text for details}
\label{zn1500testee}
\end{figure}

The total HFB energy per particle without $V_{p-e}$ is $\left.\frac{E}{A}\right|_{no\; V_{p-e}}=5.883$ MeV, while if we switch it on we obtain  $\left.\frac{E}{A}\right|_{tot}=5.768$ MeV, leading to a difference
of $\approx 116$ keV per particle.
As the kinetic energy of the electrons is much higher than their Coulomb interaction with the protons,
such a small energy difference was expected.
Nevertheless, as neighbour $(Z,\rho_b)$ configurations around the energy minima are separated 
by energies of the order of a few keV~\cite{pea12}, it is essential to calculate the total energy of the system 
with the highest precision possible. The small contribution of the $V_{p-e}$ potential is then important.
The effect of $V_{p-e}$ on the density profiles is negligible, as it is shown in Fig.\ref{zn1500testee}.

Since $V_{p-e}$ explicitly depends on the value of $R_{WS}$, one needs to minimize the cluster configurations in a three dimensional space $E_{tot}(R_{WS},Z,\rho_b)$. This is the procedure followed for example in Ref.~\cite{gri11}. This is computationally too expensive and given the very small effects of $V_{p-e}$ on the total energy, we have decided to treat it as a perturbation.
In this case, the total energy per particle is the sum of the total energy without the proton-electron
interaction plus a perturbative estimate of such interaction

\begin{eqnarray}
  \left.\frac{E}{A}\right|_{tot\; perturb.}=\left.\frac{E}{A}\right|_{tot,\; no \; V_{p-e}}+\frac{1}{A}E_{pe}(R_{WS})\;,
\end{eqnarray}

\noindent where $E_{pe}$ reads

\begin{eqnarray}
E_{pe}(R_{WS})=-\frac{3}{2}\frac{Z N_e e^2}{R_{WS}}+2\pi \frac{e^2 N_e}{R_{WS}^3} \int_0^{R_{WS}}dr \rho_p(r) r^4\,
\end{eqnarray}

\noindent  and $N_e$ is the number of electrons.
For the $^{1500}$Zr cell, we obtain $\left.\frac{E}{A}\right|_{tot\; perturb.}=5.770$ MeV, leading to an error 
$\sigma_{pe}=2$ keV with respect to the fully self consistent result. We include the proton-electron contribution perturbatively for all our calculations,
without significantly increasing the error on the total HFB energy.

As the proton state energies are shifted to lower values, the proton chemical potential is also shifted and
it can be calculated as

\begin{eqnarray}\label{eq:mup}
\mu_p=\varepsilon_{F,p}+\frac{d E_{pe}}{dZ}\;.
\end{eqnarray}  

\noindent where $\varepsilon_{F,p}$ is the HFB proton chemical potential without the contribution of the $V_{p-e}$ potential.
The value we obtain with this procedure is within less than 100 keV from the value of the fully self-consistent calculation, $i.e.$ including $V_{p-e}$ into the HFB calculation.

\subsection{$\beta$-equilibrium condition}\label{sect:beta_eq}

Each WS cell is characterized by containing a given fraction of protons, electrons and neutrons in $\beta$-equilibrium.
For each cluster composition and for each baryonic density, we thus need to calculate the configuration at $\beta$-equilibrium and thus determining the size $R_{WS}$ of the WS cell and its neutron density. This means that
we have to impose Eq.\ref{eq:beta_equilibrium_equal}. 
See Ref.~\cite{gri11} for more details.

Eq.\ref{eq:beta_equilibrium_equal} is solved by varying $R_{WS}$ for a fixed baryonic density and fixed cluster configurations. The calculations are very fast and the equation is solved exactly.

It is worth noticing that any error introduced on the proton chemical potential would propagate to 
the total energy of the WS and also to its radius.
By performing several tests, we estimated that a perturbative chemical potential introduces an additional error
$\sigma_{\beta} \leq 3$ keV on the total energy per particle.
This error has a strong density dependence, since the additional term in Eq.\ref{eq:mup} strongly depends on the curvature 
of the potential and thus on the size of the cell and the average baryonic density. 
As in Sect.~\ref{ep_energy}, we have used the WS  $^{1500}$Zr cell from Ref.~\cite{Negele1973} since we expect the contribution of $V_{p-e}$ to be more important and thus the perturbative treatment to be less efficient for this cell.
From this analysis we have obtained a conservative estimate of 3 keV per particle on the total energy of the WS cell. A better treatment of such a term in the future could lead to a reduction of the error, but at the moment it is not really important since it is of the same order of magnitude of the other sources of error.

We also estimated the error on the determination of the WS cell radius to be of $\pm0.2$ fm due to the inaccuracy on the determination of $\mu_p$. This  value is compatible with the mesh size adopted in Ref.~\cite{gri11} to scan the different $\beta$-equilibrium conditions, but with much lower computational cost.
%

%
\subsection{Summary of the sources of error}\label{summary_errors}

In the previous Sections, we estimated the different sources of error on the HFB total energy. We have essentially three major sources:
\begin{enumerate}[label=(\roman*)]
\item $\sigma_{box} \approx 1$ keV, arising from the state discretization.
\item $\sigma_\beta \approx 3$ keV, due to a non exact fulfilment of $\beta$-equilibrium condition due to an inaccuracy in determining the proton chemical potential.
\item $\sigma_{pe} \approx 2$ keV, due to a perturbative treatment of the proton-electron potential. 
\end{enumerate}

\noindent Other possible sources of error are related to the numerical integrations done in the numerical code used to solve HFB equations. For this purpose we have implemented the Gauss-Legendre (GL) quadrature to improve the accuracy of the numerical integrals. We have used 400 GL points to minimise all sources of numerical noise.
The derivatives are performed analytically using the properties of Bessel functions~\cite{abramowitz1964handbook}. In doing that, we have been able to reduce the numerical noise to the eV level, thus negligible compared to other sources of error.

The total error per particle on the HFB energy reads

\begin{eqnarray}\label{err:HFB}
\sigma_{HFB }=\sqrt{\sigma_{box}^2+\sigma_{\beta}^2+\sigma_{pe}^2}\approx 4 \text{keV}\;.
\end{eqnarray}

This is of course just an estimate of the total error, as $\sigma_\beta$ and $\sigma_{pe}$ are not
completely independent sources of error and the sum in quadrature is not fully justified.This is the uncertainty that was given to the GPE model.

As discussed before, several authors have made use of other approximation trying to eliminate the spurious shell effects in the free neutron gas. These methods are mainly based on semi-classical approximations based on Thomas-Fermi approximations~\cite{brack1985selfconsistent}, eventually corrected by shell effects on bound states via Strutinsky method~\cite{strutinsky1967shell}.

In Ref.~\cite{pastore2016pos}, we have shown that the errors are typically one order of magnitude larger than the value estimated in Eq. \ref{err:HFB} when semi-classical methods are adopted ($i.e.$ neglecting the pairing correlations for the neutrons).
A more detailed comparison between some semi-classical methods and quantal Hartree-Fock calculations has been done in Ref.~\cite{aymard2014medium}. In this case the conclusion is that semi-classical approximations (with no Strutinsky correction) lead to a typical error of $\approx200$ keV per nucleon.

In the present article, we do not discuss the uncertainties related to the choice of the functional  and to the associated intrinsic error~\cite{toivanen2008error,gao2013propagation,haverinen2016uncertainty}.
As a word of caution, it should be mentioned that all current functionals have coupling constant that are typically adjusted on properties of finite nuclei and eventually on $ab-initio$ calculations done in the infinite medium~\cite{cha97,Gor10}.
To study the NS crust, we thus need to rely on extrapolations to very neutron rich regions that have not been included explicitly into the fit. A set of single results based on a single functional should not be used to make predictions on the property of the crust.

In Ref.~\cite{utama2016nuclear} a Bayesian neural network (BNN) formalism has been studied to improve the predictions on the structure of the outer crust of NS. We think that this method could be considered more reliable than a single extrapolation based on a single functional.
Thanks to the GPE method introduced here, the BNN formalism could be now extended to study the properties of the inner crust of a NS to assess the major uncertainties arising from different functionals. 


%
\section{Cluster structure}\label{cluster_structure}

In this section, we present our results combining the HFB calculations and the GPE results. As anticipated in Sec.\ref{the_method}, we assume that the average baryonic density can vary in the interval $\rho_b\in[0.0004,0.02]$ fm$^{-3}$.

\begin{figure*}[!h]
\begin{center}
\includegraphics[width=0.340\textwidth,angle=-90]{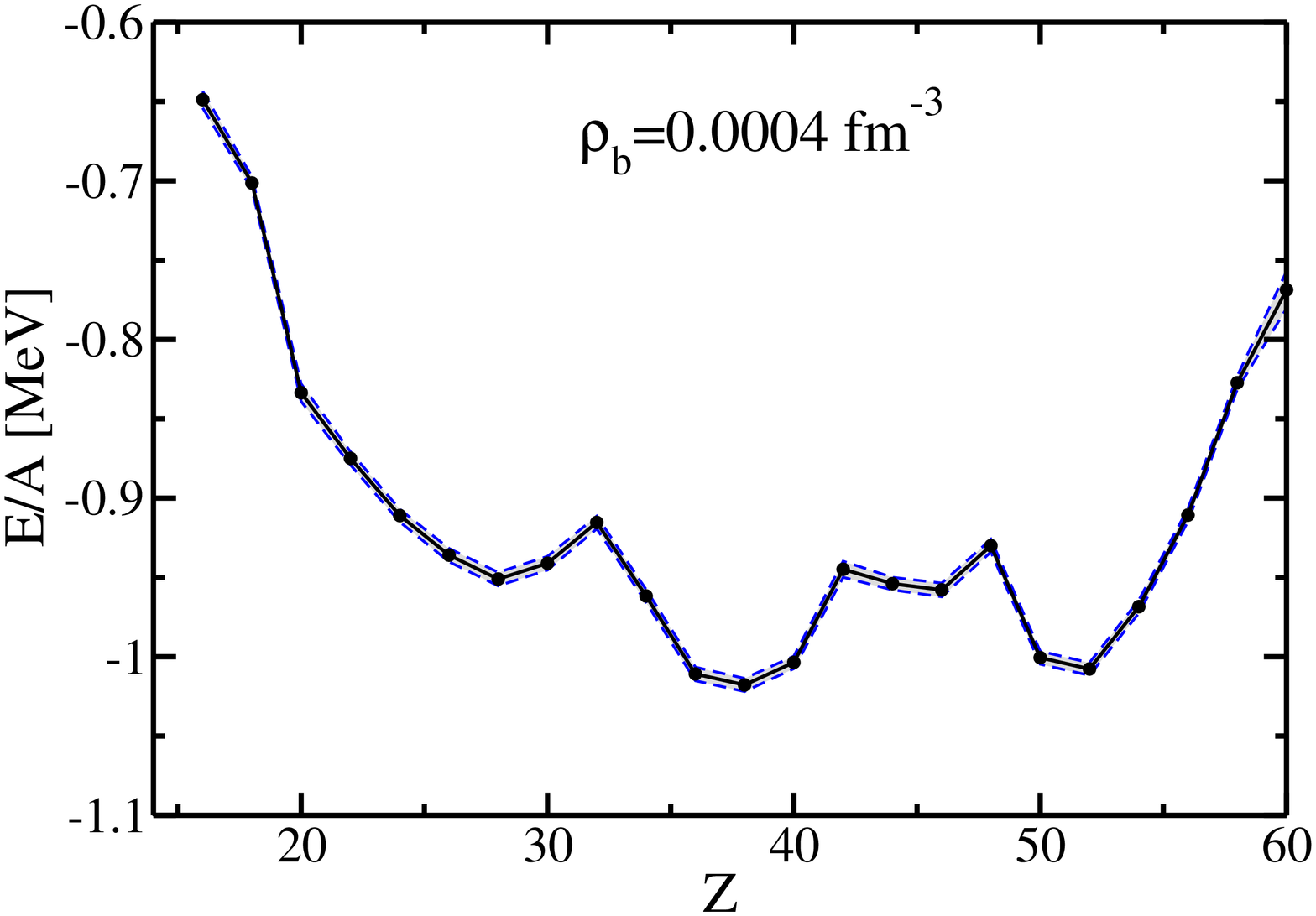}
\includegraphics[width=0.340\textwidth,angle=-90]{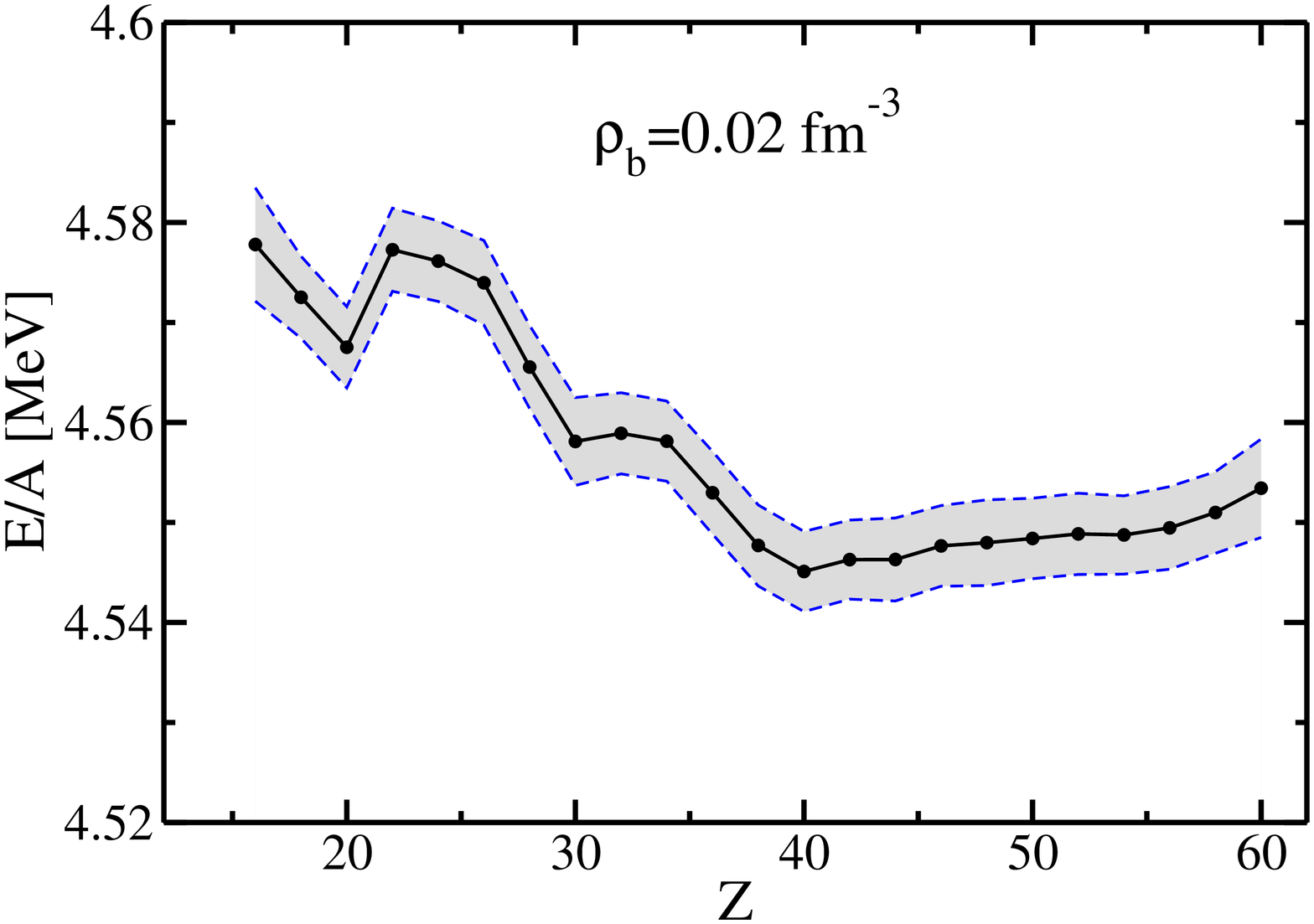}
\end{center}
\caption{(Colors online) Total energy per particle as a function of the number of protons in the cluster at the center of the WS cell. The solid black lines correspond to the curve extracted using the GPE and the dashed are between the dashed line represent the 90\% confidence interval. See text for details.}
\label{Minima:strong}
\end{figure*}

In Fig.\ref{Minima:strong}, we show the total energy per particle as a function of the proton charge for the two limit values of this interval. The solid lines represent the result obtained using the GPE while the dashed lines represent the 90\% confidence intervals.
As already noticed by other authors~\cite{gri11,pea12,sha15}, for a given value of $\rho_b$ there are typically several competing local minima, usually around the proton shell closures. Such an effect was also observed in Ref.~\cite{Negele1973}. 
In particular, we notice two well pronounced minima around Z=40 and Z=50 in the low density regions. In Ref.~\cite{sha15}, where shell effects are not included, this feature is not present and the proton clusters are typically smaller $\text{Z}\in [24-36]$.

In both panels of Fig.\ref{Minima:strong}, we see that although the absolute energy scales are quite different, the energy minima are usually quite close to each other and they differ by a few keV. The energy differences between the minima decreases as the average baryonic density increases~\cite{pea12}.
On the right panel of Fig.\ref{Minima:strong}, we observe that although the absolute minimum turns out to be Z=40, all values of Z between 40 and 54 are within the $\sigma_{HFB}$ uncertainty of Eq.~\ref{err:HFB}.
 
\begin{figure}[!h]
\begin{center}
\includegraphics[width=0.40\textwidth,angle=0]{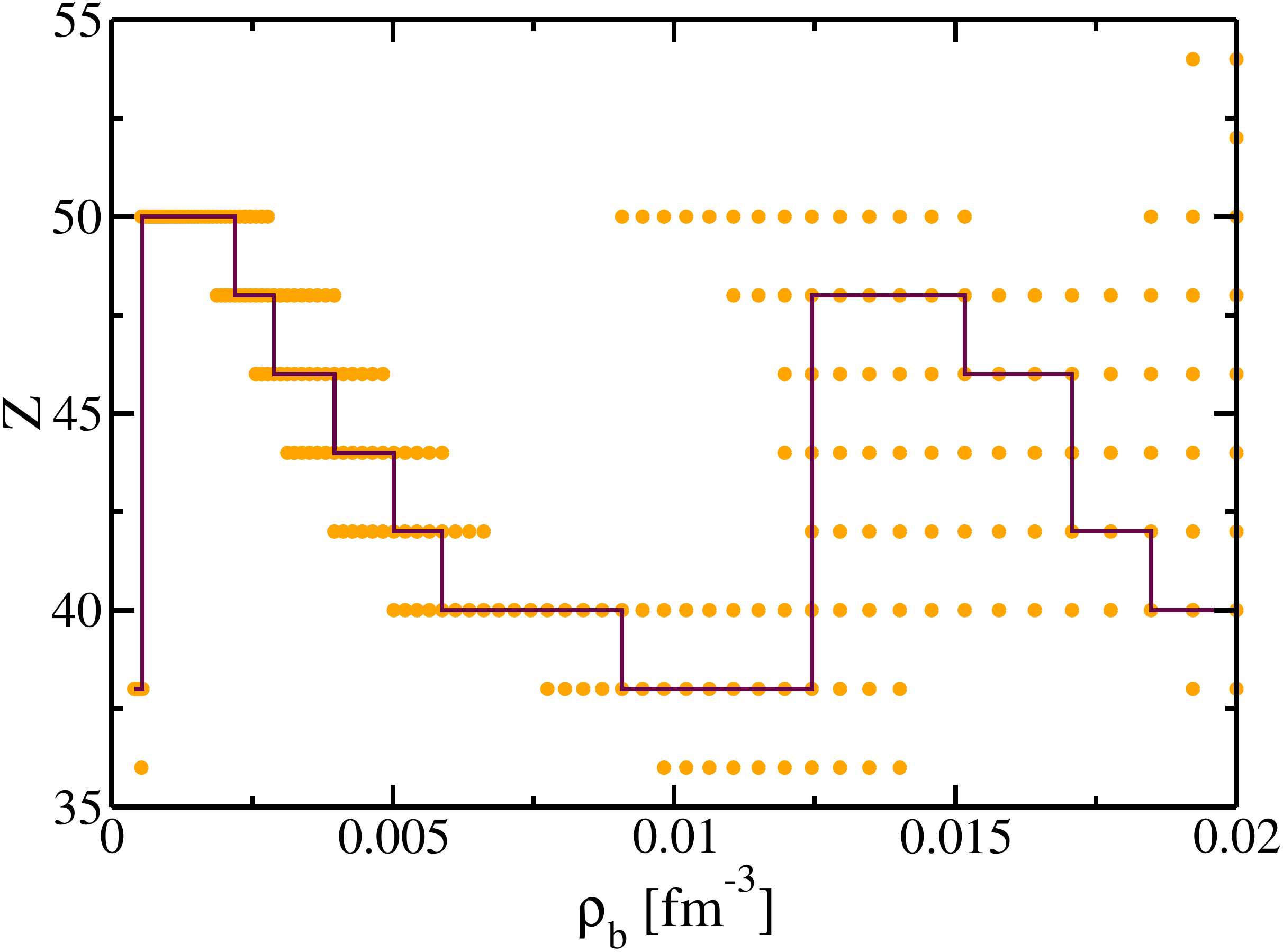}
\end{center}
\caption{(Colors online) The ($Z, \rho_b$) configurations corresponding to the absolute energy minima. The solid line represent the value of the proton charge as a function of the average baryonic density that leads to minimal energy configuration. The dots on the figure represents the other possible configurations that have a total energy per particle that differs less than the estimated error bar respect to the absolute minimum. See text for details.}
\label{minimum_energy_path}
\end{figure}

In Fig.\ref{minimum_energy_path}, we present the position of the absolute minima extracted from our calculations (solid line) for the region of density considered in this work.
In the low density region, we observe a very well pronounced minimum at Z=38. This value is compatible  with the typical drip-line nuclei found in the outer crust~\cite{roca2008impact,pearson2011properties,sha15}.
At $\rho_b=0.0006$ fm$^{-3}$, we observe a transition from Z=38 to Z=50. A similar transition was also predicted in Ref.~\cite{gri11} at similar values of the density where the same functionals has been used.
Starting from $\rho_b=0.002$ fm$^{-3}$ the cluster gradually becomes lighter and lighter, until we find again Z=38 at $\rho_b=0.01$ fm$^{-3}$. Similar behaviour was found in Ref.~\cite{gri11}, but apart from a qualitative agreement the values of Z are remarkably different. In our case we never obtain cluster smaller than Z=38, while in Ref.~\cite{gri11}, Ca isotopes becomes favourable at high density.

Another interesting feature of our calculation is that at $\rho_b=0.012$ fm$^{-3}$ there is another jump from Z=38 to Z=50 followed by a gradual decrease to find finally Z=40 at $\rho_b=0.02$ fm$^{-3}$.

These results should be taken with a grain of salt. Given the total error $\sigma_{HFB} \approx 4$keV per particle, as discussed in Sec.~\ref{summary_errors}, we have added on Fig.\ref{minimum_energy_path} the other possible values of Z falling within this error bar.
We observe that in the interval $\rho_b\in[0.0004,0.01]$ fm$^{-3}$ our results are quite robust. Beyond $\rho_b=0.01$ fm$^{-3}$ all values of proton number within the range $38\le Z \le 50$ are possible energy minima, given the numerical accuracy of our calculations. For such a reason, we have decided not to present the results beyond $\rho_b=0.02$ fm$^{-3}$ since our numerical method is not accurate enough to give a sensible prediction of the cluster composition of the star in this region.

\begin{figure}[!h]
\begin{center}
\includegraphics[width=0.340\textwidth,angle=-90]{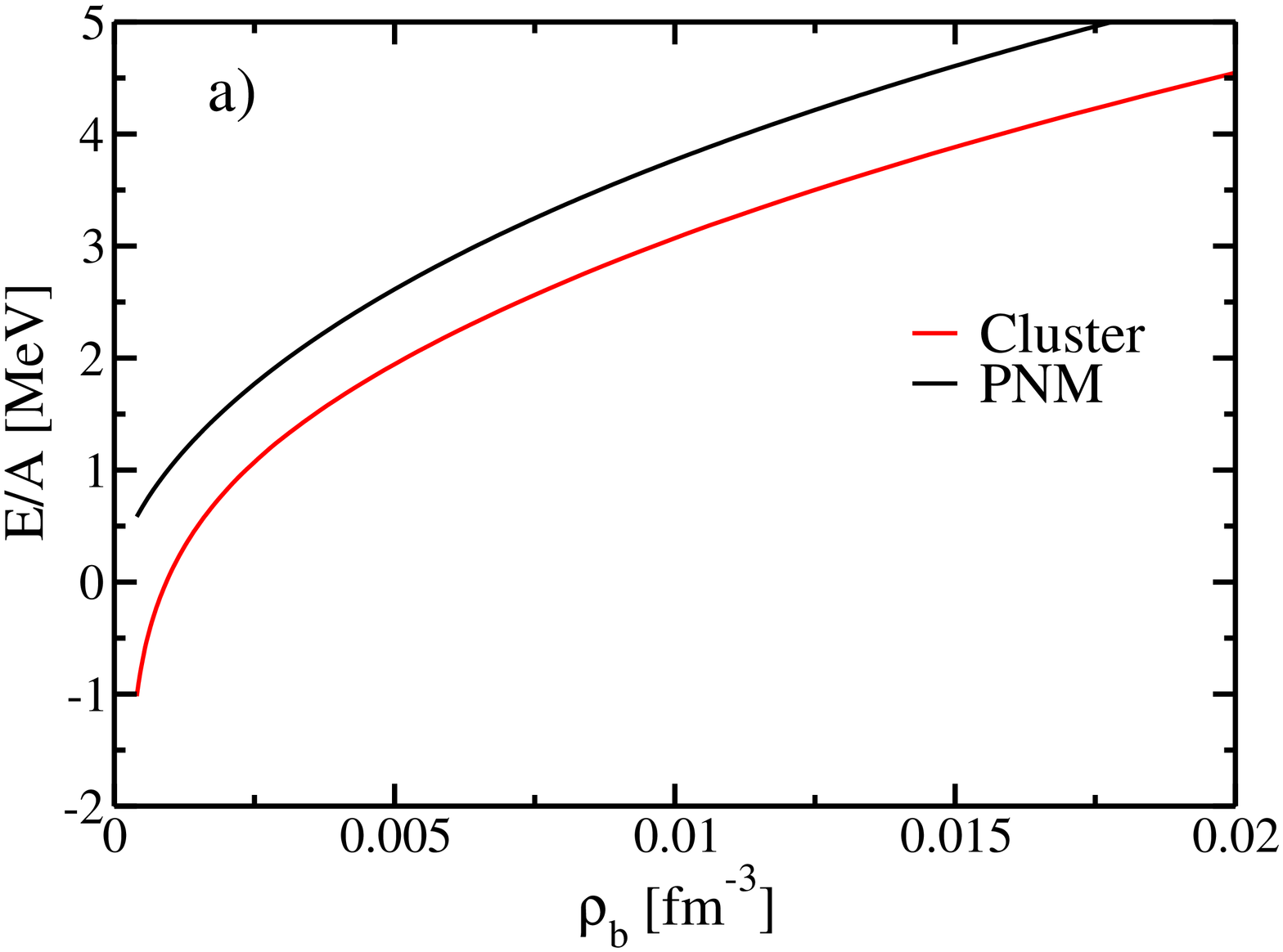}
\includegraphics[width=0.340\textwidth,angle=-90]{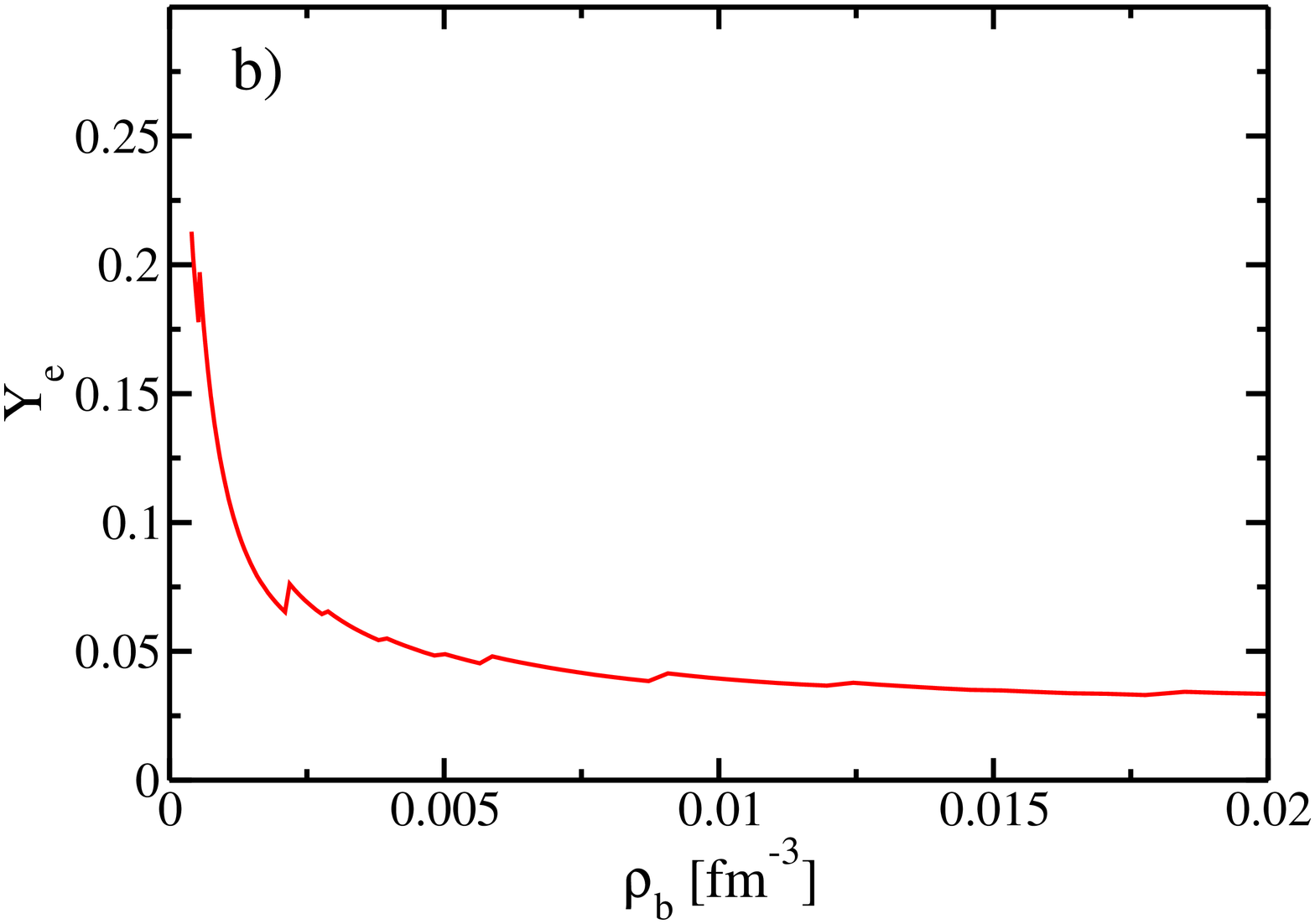}
\end{center}
\caption{(Colors online) On panel (a), the energy per particle in the case of pure neutron matter and in the case of clustered WS cells as a function of the average baryonic density. On  panel (b), the proton fraction obtained in our calculations. See text for details.}
\label{esura}
\end{figure}

We can calculate the total energy per particle for each of the WS cells at the energy minima. The result is shown in Fig.\ref{esura}(a),
together with the energy per particle obtained for a free neutron gas. We see that at these densities it is energetically favourable to form clusters.
Another important quantity we can calculate is the proton fraction $Y=Z/A$ of each WS cell at the energy minima. 
We present the results in Fig.\ref{esura}(b).
This quantity  is less sensitive to the details of the cluster composition as already observed in Ref.~\cite{gulminelli2015unified}, since the main contribution comes from the energetics of the electrons.

\section{Conclusions}\label{sec:concl}


\noindent We have performed a detailed study to determine the proton component 
of the inner crust of a neutron star, solving the fully self-consistent Hartree-Fock-Bogoliubov equations
in the WS approximation.
We have developed a very accurate numerical procedure to reduce all possible sources of numerical noise. In particular, we have studied the effects of the box size to minimize the spurious shell-effects on the external neutron gas of the WS cell. By using very large boxes and thus large number of basis states, we have been able to reduce this effect to the order of 1 keV per particle.

To further reduce the computational cost of our procedure, we have treated the proton-electron potential in a perturbative way. At the price of adding extra 2-3 keV per particle, we have thus reduced the dimension of the configuration space 
that one needs to explore to find the energy minima.

Although this approximation strongly reduces the required CPU time, a complete exploration of the $(Z,\rho_b)$ surface is still too demanding using only HFB methods.
We have thus explored, for the very first time, the use of a Gaussian  Process Emulator combined with the HFB results.
We have demonstrated that the GPE  method can reduce by a factor of 10 the number of required HFB 
calculations that are needed to determine the cluster configurations of the inner crust. All the extrapolated results obtained with GPE fall within the estimated error bar $\sigma_{HFB}$.

In the future, we thus plan to apply systematically the GPE techniques to the determination of the properties of the inner crust. Although the execution time would still be larger than a semi-classical method~\cite{pea12}, it allows us to treat at the same time shell effects and pairing correlations.
As a final remark, we stress that the GPE has been developed independently by the HFB codes, thus it will be possible to test it in future applications  using different simulators based on other approaches, as for example using more solvers based on band-theory ~\cite{chamel2005band}.

\section*{Acknowledgments}
Calculations were performed by using HPC resources from GENCI-TGCC (Contracts No. 2015-057392 and No. 2016-057392) and the DiRAC Data Analytic system at the University of Cambridge (under BIS National E-infrastructure capital Grant No. ST/J005673/1, and STFC Grants No. ST/H008586/1 and No. ST/K00333X/1).


\bibliography{biblio}

\begin{thebibliography}{87}
\expandafter\ifx\csname natexlab\endcsname\relax\def\natexlab#1{#1}\fi
\expandafter\ifx\csname bibnamefont\endcsname\relax
  \def\bibnamefont#1{#1}\fi
\expandafter\ifx\csname bibfnamefont\endcsname\relax
  \def\bibfnamefont#1{#1}\fi
\expandafter\ifx\csname citenamefont\endcsname\relax
  \def\citenamefont#1{#1}\fi
\expandafter\ifx\csname url\endcsname\relax
  \def\url#1{\texttt{#1}}\fi
\expandafter\ifx\csname urlprefix\endcsname\relax\def\urlprefix{URL }\fi
\providecommand{\bibinfo}[2]{#2}
\providecommand{\eprint}[2][]{\url{#2}}

\bibitem[{\citenamefont{Hewish et~al.}(1968)\citenamefont{Hewish, Bell,
  Pilkington, Scott, and Collins}}]{hew68}
\bibinfo{author}{\bibfnamefont{A.}~\bibnamefont{Hewish}},
  \bibinfo{author}{\bibfnamefont{S.~J.} \bibnamefont{Bell}},
  \bibinfo{author}{\bibfnamefont{J.}~\bibnamefont{Pilkington}},
  \bibinfo{author}{\bibfnamefont{P.~F.} \bibnamefont{Scott}}, \bibnamefont{and}
  \bibinfo{author}{\bibfnamefont{R.~A.} \bibnamefont{Collins}},
  \bibinfo{journal}{Nature} \textbf{\bibinfo{volume}{217}},
  \bibinfo{pages}{709} (\bibinfo{year}{1968}).

\bibitem[{\citenamefont{Hulse}(1994)}]{hul94}
\bibinfo{author}{\bibfnamefont{R.~A.} \bibnamefont{Hulse}},
  \bibinfo{journal}{Reviews of Modern Physics} \textbf{\bibinfo{volume}{66}},
  \bibinfo{pages}{699} (\bibinfo{year}{1994}).

\bibitem[{\citenamefont{Manchester et~al.}(2001)\citenamefont{Manchester, Lyne,
  Camilo, Bell, Kaspi, D'Amico, McKay, Crawford, Stairs, Possenti
  et~al.}}]{man01}
\bibinfo{author}{\bibfnamefont{R.~N.} \bibnamefont{Manchester}},
  \bibinfo{author}{\bibfnamefont{A.~G.} \bibnamefont{Lyne}},
  \bibinfo{author}{\bibfnamefont{F.}~\bibnamefont{Camilo}},
  \bibinfo{author}{\bibfnamefont{J.}~\bibnamefont{Bell}},
  \bibinfo{author}{\bibfnamefont{V.}~\bibnamefont{Kaspi}},
  \bibinfo{author}{\bibfnamefont{N.}~\bibnamefont{D'Amico}},
  \bibinfo{author}{\bibfnamefont{N.}~\bibnamefont{McKay}},
  \bibinfo{author}{\bibfnamefont{F.}~\bibnamefont{Crawford}},
  \bibinfo{author}{\bibfnamefont{I.}~\bibnamefont{Stairs}},
  \bibinfo{author}{\bibfnamefont{A.}~\bibnamefont{Possenti}},
  \bibnamefont{et~al.}, \bibinfo{journal}{Monthly Notices of the Royal
  Astronomical Society} \textbf{\bibinfo{volume}{328}}, \bibinfo{pages}{17}
  (\bibinfo{year}{2001}).

\bibitem[{\citenamefont{Manchester}(2004)}]{manchester2004pulsars}
\bibinfo{author}{\bibfnamefont{R.~N.} \bibnamefont{Manchester}},
  \bibinfo{journal}{Science} \textbf{\bibinfo{volume}{304}},
  \bibinfo{pages}{542} (\bibinfo{year}{2004}).

\bibitem[{\citenamefont{Lattimer and Prakash}(2004)}]{lattimer2004neutronstars}
\bibinfo{author}{\bibfnamefont{J.~M.} \bibnamefont{Lattimer}} \bibnamefont{and}
  \bibinfo{author}{\bibfnamefont{M.}~\bibnamefont{Prakash}},
  \bibinfo{journal}{Science} \textbf{\bibinfo{volume}{304}},
  \bibinfo{pages}{536} (\bibinfo{year}{2004}).

\bibitem[{\citenamefont{Michel}(1991)}]{mic91}
\bibinfo{author}{\bibfnamefont{F.~C.} \bibnamefont{Michel}},
  \emph{\bibinfo{title}{Theory of neutron star magnetospheres}}
  (\bibinfo{publisher}{University of Chicago Press}, \bibinfo{year}{1991}).

\bibitem[{\citenamefont{Haensel et~al.}(2007)\citenamefont{Haensel, Potekhin,
  and Yakovlev}}]{hae07}
\bibinfo{author}{\bibfnamefont{P.}~\bibnamefont{Haensel}},
  \bibinfo{author}{\bibfnamefont{A.~Y.} \bibnamefont{Potekhin}},
  \bibnamefont{and} \bibinfo{author}{\bibfnamefont{D.~G.}
  \bibnamefont{Yakovlev}}, \emph{\bibinfo{title}{Neutron stars 1: Equation of
  state and structure}}, vol. \bibinfo{volume}{326}
  (\bibinfo{publisher}{Springer Science \& Business Media},
  \bibinfo{year}{2007}).

\bibitem[{\citenamefont{Reisenegger and
  Zepeda}(2016)}]{reisenegger2016ordermagnitude}
\bibinfo{author}{\bibfnamefont{A.}~\bibnamefont{Reisenegger}} \bibnamefont{and}
  \bibinfo{author}{\bibfnamefont{F.~S.} \bibnamefont{Zepeda}},
  \bibinfo{journal}{Eur. Phys. J A} \textbf{\bibinfo{volume}{52}},
  \bibinfo{pages}{52} (\bibinfo{year}{2016}).

\bibitem[{\citenamefont{Lattimer}(2005)}]{lattimer2005ultimate}
\bibinfo{author}{\bibfnamefont{J.~M.} \bibnamefont{Lattimer}},
  \bibinfo{journal}{Phys. Rev. Lett.} \textbf{\bibinfo{volume}{94}},
  \bibinfo{pages}{111101} (\bibinfo{year}{2005}).

\bibitem[{\citenamefont{Demorest et~al.}(2010)\citenamefont{Demorest, Pennucci,
  Ransom, Roberts, and Hessels}}]{demorest2010twosolarmasses}
\bibinfo{author}{\bibfnamefont{P.~B.} \bibnamefont{Demorest}},
  \bibinfo{author}{\bibfnamefont{T.}~\bibnamefont{Pennucci}},
  \bibinfo{author}{\bibfnamefont{S.~M.} \bibnamefont{Ransom}},
  \bibinfo{author}{\bibfnamefont{M.~S.~E.} \bibnamefont{Roberts}},
  \bibnamefont{and} \bibinfo{author}{\bibfnamefont{J.~W.~T.}
  \bibnamefont{Hessels}}, \bibinfo{journal}{Nature}
  \textbf{\bibinfo{volume}{467}}, \bibinfo{pages}{1081} (\bibinfo{year}{2010}).

\bibitem[{\citenamefont{Antoniadis et~al.}(2013)\citenamefont{Antoniadis,
  Freire, Wex, Tauris, Lynch, van Kerkwijk, Kramer, Bassa, Dhillon, Driebe
  et~al.}}]{antoniadis2013massive}
\bibinfo{author}{\bibfnamefont{J.}~\bibnamefont{Antoniadis}},
  \bibinfo{author}{\bibfnamefont{P.~C.} \bibnamefont{Freire}},
  \bibinfo{author}{\bibfnamefont{N.}~\bibnamefont{Wex}},
  \bibinfo{author}{\bibfnamefont{T.~M.} \bibnamefont{Tauris}},
  \bibinfo{author}{\bibfnamefont{R.~S.} \bibnamefont{Lynch}},
  \bibinfo{author}{\bibfnamefont{M.~H.} \bibnamefont{van Kerkwijk}},
  \bibinfo{author}{\bibfnamefont{M.}~\bibnamefont{Kramer}},
  \bibinfo{author}{\bibfnamefont{C.}~\bibnamefont{Bassa}},
  \bibinfo{author}{\bibfnamefont{V.~S.} \bibnamefont{Dhillon}},
  \bibinfo{author}{\bibfnamefont{T.}~\bibnamefont{Driebe}},
  \bibnamefont{et~al.}, \bibinfo{journal}{Science}
  \textbf{\bibinfo{volume}{340}}, \bibinfo{pages}{1233232}
  (\bibinfo{year}{2013}).

\bibitem[{\citenamefont{Fonseca et~al.}(2016)\citenamefont{Fonseca, Pennucci,
  Ellis, Stairs, Nice, Ransom, Demorest, Arzoumanian, Crowter, Dolch
  et~al.}}]{fonseca2016nanograv}
\bibinfo{author}{\bibfnamefont{E.}~\bibnamefont{Fonseca}},
  \bibinfo{author}{\bibfnamefont{T.~T.} \bibnamefont{Pennucci}},
  \bibinfo{author}{\bibfnamefont{J.~A.} \bibnamefont{Ellis}},
  \bibinfo{author}{\bibfnamefont{I.~H.} \bibnamefont{Stairs}},
  \bibinfo{author}{\bibfnamefont{D.~J.} \bibnamefont{Nice}},
  \bibinfo{author}{\bibfnamefont{S.~M.} \bibnamefont{Ransom}},
  \bibinfo{author}{\bibfnamefont{P.~B.} \bibnamefont{Demorest}},
  \bibinfo{author}{\bibfnamefont{Z.}~\bibnamefont{Arzoumanian}},
  \bibinfo{author}{\bibfnamefont{K.}~\bibnamefont{Crowter}},
  \bibinfo{author}{\bibfnamefont{T.}~\bibnamefont{Dolch}},
  \bibnamefont{et~al.}, \bibinfo{journal}{The Astrophysical Journal}
  \textbf{\bibinfo{volume}{832}}, \bibinfo{pages}{167} (\bibinfo{year}{2016}).

\bibitem[{\citenamefont{Lattimer and Prakash}(2007)}]{lattimer2007neutronstars}
\bibinfo{author}{\bibfnamefont{J.~M.} \bibnamefont{Lattimer}} \bibnamefont{and}
  \bibinfo{author}{\bibfnamefont{M.}~\bibnamefont{Prakash}},
  \bibinfo{journal}{Phys. Rep.} \textbf{\bibinfo{volume}{442}},
  \bibinfo{pages}{109} (\bibinfo{year}{2007}).

\bibitem[{\citenamefont{Glendenning and
  Schaffner-Bielich}(1998)}]{glendenning1998kaons}
\bibinfo{author}{\bibfnamefont{N.~K.} \bibnamefont{Glendenning}}
  \bibnamefont{and}
  \bibinfo{author}{\bibfnamefont{J.}~\bibnamefont{Schaffner-Bielich}},
  \bibinfo{journal}{Phys. Rev. Lett.} \textbf{\bibinfo{volume}{81}},
  \bibinfo{pages}{4564} (\bibinfo{year}{1998}).

\bibitem[{\citenamefont{Lackey et~al.}(2006)\citenamefont{Lackey, Nayyar, and
  Owen}}]{lackey2006hyperons}
\bibinfo{author}{\bibfnamefont{B.~D.} \bibnamefont{Lackey}},
  \bibinfo{author}{\bibfnamefont{M.}~\bibnamefont{Nayyar}}, \bibnamefont{and}
  \bibinfo{author}{\bibfnamefont{B.~J.} \bibnamefont{Owen}},
  \bibinfo{journal}{Phys. Rev. D} \textbf{\bibinfo{volume}{73}},
  \bibinfo{pages}{024021} (\bibinfo{year}{2006}).

\bibitem[{\citenamefont{Vida\~na et~al.}(2000)\citenamefont{Vida\~na, Polls,
  Ramos, Engvik, and Hjorth-Jensen}}]{vid00}
\bibinfo{author}{\bibfnamefont{I.}~\bibnamefont{Vida\~na}},
  \bibinfo{author}{\bibfnamefont{A.}~\bibnamefont{Polls}},
  \bibinfo{author}{\bibfnamefont{A.}~\bibnamefont{Ramos}},
  \bibinfo{author}{\bibfnamefont{L.}~\bibnamefont{Engvik}}, \bibnamefont{and}
  \bibinfo{author}{\bibfnamefont{M.}~\bibnamefont{Hjorth-Jensen}},
  \bibinfo{journal}{Phys. Rev. C} \textbf{\bibinfo{volume}{62}},
  \bibinfo{pages}{035801} (\bibinfo{year}{2000}),
  \urlprefix\url{https://link.aps.org/doi/10.1103/PhysRevC.62.035801}.

\bibitem[{\citenamefont{Astashenok et~al.}(2014)\citenamefont{Astashenok,
  Capozziello, and Odintsov}}]{ast14}
\bibinfo{author}{\bibfnamefont{A.~V.} \bibnamefont{Astashenok}},
  \bibinfo{author}{\bibfnamefont{S.}~\bibnamefont{Capozziello}},
  \bibnamefont{and} \bibinfo{author}{\bibfnamefont{S.~D.}
  \bibnamefont{Odintsov}}, \bibinfo{journal}{Physical Review D}
  \textbf{\bibinfo{volume}{89}}, \bibinfo{pages}{103509}
  (\bibinfo{year}{2014}).

\bibitem[{\citenamefont{Lonardoni et~al.}(2015)\citenamefont{Lonardoni, Lovato,
  Gandolfi, and Pederiva}}]{lon15}
\bibinfo{author}{\bibfnamefont{D.}~\bibnamefont{Lonardoni}},
  \bibinfo{author}{\bibfnamefont{A.}~\bibnamefont{Lovato}},
  \bibinfo{author}{\bibfnamefont{S.}~\bibnamefont{Gandolfi}}, \bibnamefont{and}
  \bibinfo{author}{\bibfnamefont{F.}~\bibnamefont{Pederiva}},
  \bibinfo{journal}{Physical review letters} \textbf{\bibinfo{volume}{114}},
  \bibinfo{pages}{092301} (\bibinfo{year}{2015}).

\bibitem[{\citenamefont{Chamel and Haensel}(2008)}]{cha08}
\bibinfo{author}{\bibfnamefont{N.}~\bibnamefont{Chamel}} \bibnamefont{and}
  \bibinfo{author}{\bibfnamefont{P.}~\bibnamefont{Haensel}},
  \bibinfo{journal}{Living Rev.Rel.} \textbf{\bibinfo{volume}{11}},
  \bibinfo{pages}{10} (\bibinfo{year}{2008}).

\bibitem[{\citenamefont{Chatterjee and Vida{\~n}a}(2016)}]{cha16}
\bibinfo{author}{\bibfnamefont{D.}~\bibnamefont{Chatterjee}} \bibnamefont{and}
  \bibinfo{author}{\bibfnamefont{I.}~\bibnamefont{Vida{\~n}a}},
  \bibinfo{journal}{The European Physical Journal A}
  \textbf{\bibinfo{volume}{52}}, \bibinfo{pages}{1} (\bibinfo{year}{2016}).

\bibitem[{\citenamefont{Douchin and Haensel}(2000)}]{douchin2000inner}
\bibinfo{author}{\bibfnamefont{F.}~\bibnamefont{Douchin}} \bibnamefont{and}
  \bibinfo{author}{\bibfnamefont{P.}~\bibnamefont{Haensel}},
  \bibinfo{journal}{Physics Letters B} \textbf{\bibinfo{volume}{485}},
  \bibinfo{pages}{107} (\bibinfo{year}{2000}).

\bibitem[{\citenamefont{Baldo et~al.}(1997)\citenamefont{Baldo, Bombaci, and
  Burgio}}]{baldo1997microscopic}
\bibinfo{author}{\bibfnamefont{M.}~\bibnamefont{Baldo}},
  \bibinfo{author}{\bibfnamefont{I.}~\bibnamefont{Bombaci}}, \bibnamefont{and}
  \bibinfo{author}{\bibfnamefont{G.}~\bibnamefont{Burgio}},
  \bibinfo{journal}{Astronomy and Astrophysics} \textbf{\bibinfo{volume}{328}},
  \bibinfo{pages}{274} (\bibinfo{year}{1997}).

\bibitem[{\citenamefont{Chen et~al.}(2005)\citenamefont{Chen, Ko, and
  Li}}]{chen2005determination}
\bibinfo{author}{\bibfnamefont{L.-W.} \bibnamefont{Chen}},
  \bibinfo{author}{\bibfnamefont{C.~M.} \bibnamefont{Ko}}, \bibnamefont{and}
  \bibinfo{author}{\bibfnamefont{B.-A.} \bibnamefont{Li}},
  \bibinfo{journal}{Physical review letters} \textbf{\bibinfo{volume}{94}},
  \bibinfo{pages}{032701} (\bibinfo{year}{2005}).

\bibitem[{\citenamefont{Centelles et~al.}(2009)\citenamefont{Centelles,
  Roca-Maza, Vinas, and Warda}}]{centelles2009nuclear}
\bibinfo{author}{\bibfnamefont{M.}~\bibnamefont{Centelles}},
  \bibinfo{author}{\bibfnamefont{X.}~\bibnamefont{Roca-Maza}},
  \bibinfo{author}{\bibfnamefont{X.}~\bibnamefont{Vinas}}, \bibnamefont{and}
  \bibinfo{author}{\bibfnamefont{M.}~\bibnamefont{Warda}},
  \bibinfo{journal}{Physical review letters} \textbf{\bibinfo{volume}{102}},
  \bibinfo{pages}{122502} (\bibinfo{year}{2009}).

\bibitem[{\citenamefont{Gandolfi et~al.}(2012)\citenamefont{Gandolfi, Carlson,
  and Reddy}}]{gandolfi2012maximum}
\bibinfo{author}{\bibfnamefont{S.}~\bibnamefont{Gandolfi}},
  \bibinfo{author}{\bibfnamefont{J.}~\bibnamefont{Carlson}}, \bibnamefont{and}
  \bibinfo{author}{\bibfnamefont{S.}~\bibnamefont{Reddy}},
  \bibinfo{journal}{Physical Review C} \textbf{\bibinfo{volume}{85}},
  \bibinfo{pages}{032801} (\bibinfo{year}{2012}).

\bibitem[{\citenamefont{Hebeler and Schwenk}(2014)}]{hebeler2014symmetry}
\bibinfo{author}{\bibfnamefont{K.}~\bibnamefont{Hebeler}} \bibnamefont{and}
  \bibinfo{author}{\bibfnamefont{A.}~\bibnamefont{Schwenk}},
  \bibinfo{journal}{The European Physical Journal A}
  \textbf{\bibinfo{volume}{50}}, \bibinfo{pages}{11} (\bibinfo{year}{2014}).

\bibitem[{\citenamefont{Ducoin et~al.}(2008)\citenamefont{Ducoin,
  Provid{\^e}ncia, Santos, Brito, and Chomaz}}]{ducoin2008cluster}
\bibinfo{author}{\bibfnamefont{C.}~\bibnamefont{Ducoin}},
  \bibinfo{author}{\bibfnamefont{C.}~\bibnamefont{Provid{\^e}ncia}},
  \bibinfo{author}{\bibfnamefont{A.~M.} \bibnamefont{Santos}},
  \bibinfo{author}{\bibfnamefont{L.}~\bibnamefont{Brito}}, \bibnamefont{and}
  \bibinfo{author}{\bibfnamefont{P.}~\bibnamefont{Chomaz}},
  \bibinfo{journal}{Physical Review C} \textbf{\bibinfo{volume}{78}},
  \bibinfo{pages}{055801} (\bibinfo{year}{2008}).

\bibitem[{\citenamefont{Pethick and Ravenhall}(1995)}]{pethick1995matter}
\bibinfo{author}{\bibfnamefont{C.}~\bibnamefont{Pethick}} \bibnamefont{and}
  \bibinfo{author}{\bibfnamefont{D.}~\bibnamefont{Ravenhall}},
  \bibinfo{journal}{Annual Review of Nuclear and Particle Science}
  \textbf{\bibinfo{volume}{45}}, \bibinfo{pages}{429} (\bibinfo{year}{1995}).

\bibitem[{\citenamefont{Yakovlev and Pethick}(2004)}]{yakovlev2004neutron}
\bibinfo{author}{\bibfnamefont{D.~G.} \bibnamefont{Yakovlev}} \bibnamefont{and}
  \bibinfo{author}{\bibfnamefont{C.}~\bibnamefont{Pethick}},
  \bibinfo{journal}{Annu. Rev. Astron. Astrophys.}
  \textbf{\bibinfo{volume}{42}}, \bibinfo{pages}{169} (\bibinfo{year}{2004}).

\bibitem[{\citenamefont{Bender et~al.}(2003)\citenamefont{Bender, Heenen, and
  Reinhard}}]{bender2003self}
\bibinfo{author}{\bibfnamefont{M.}~\bibnamefont{Bender}},
  \bibinfo{author}{\bibfnamefont{P.-H.} \bibnamefont{Heenen}},
  \bibnamefont{and} \bibinfo{author}{\bibfnamefont{P.-G.}
  \bibnamefont{Reinhard}}, \bibinfo{journal}{Reviews of Modern Physics}
  \textbf{\bibinfo{volume}{75}}, \bibinfo{pages}{121} (\bibinfo{year}{2003}).

\bibitem[{\citenamefont{Goriely et~al.}(2013)\citenamefont{Goriely, Chamel, and
  Pearson}}]{gor13b}
\bibinfo{author}{\bibfnamefont{S.}~\bibnamefont{Goriely}},
  \bibinfo{author}{\bibfnamefont{N.}~\bibnamefont{Chamel}}, \bibnamefont{and}
  \bibinfo{author}{\bibfnamefont{J.~M.} \bibnamefont{Pearson}},
  \bibinfo{journal}{Phys. Rev. C} \textbf{\bibinfo{volume}{88}},
  \bibinfo{pages}{061302} (\bibinfo{year}{2013}),
  \urlprefix\url{https://link.aps.org/doi/10.1103/PhysRevC.88.061302}.

\bibitem[{\citenamefont{Negele and Vautherin}(1973)}]{Negele1973}
\bibinfo{author}{\bibfnamefont{J.~W.} \bibnamefont{Negele}} \bibnamefont{and}
  \bibinfo{author}{\bibfnamefont{D.}~\bibnamefont{Vautherin}},
  \bibinfo{journal}{Nuclear Physics A} \textbf{\bibinfo{volume}{207}},
  \bibinfo{pages}{298} (\bibinfo{year}{1973}).

\bibitem[{\citenamefont{Wigner and Seitz}(1933)}]{wig33}
\bibinfo{author}{\bibfnamefont{E.}~\bibnamefont{Wigner}} \bibnamefont{and}
  \bibinfo{author}{\bibfnamefont{F.}~\bibnamefont{Seitz}},
  \bibinfo{journal}{Phys. Rev.} \textbf{\bibinfo{volume}{43}},
  \bibinfo{pages}{804} (\bibinfo{year}{1933}).

\bibitem[{\citenamefont{Brink and Broglia}(2005)}]{brink2005nuclear}
\bibinfo{author}{\bibfnamefont{D.~M.} \bibnamefont{Brink}} \bibnamefont{and}
  \bibinfo{author}{\bibfnamefont{R.~A.} \bibnamefont{Broglia}},
  \emph{\bibinfo{title}{Nuclear Superfluidity: pairing in finite systems}},
  vol.~\bibinfo{volume}{24} (\bibinfo{publisher}{Cambridge University Press},
  \bibinfo{year}{2005}).

\bibitem[{\citenamefont{Baldo et~al.}(2005)\citenamefont{Baldo, Lombardo,
  Saperstein, and Tolokonnikov}}]{bal05}
\bibinfo{author}{\bibfnamefont{M.}~\bibnamefont{Baldo}},
  \bibinfo{author}{\bibfnamefont{U.}~\bibnamefont{Lombardo}},
  \bibinfo{author}{\bibfnamefont{E.}~\bibnamefont{Saperstein}},
  \bibnamefont{and}
  \bibinfo{author}{\bibfnamefont{S.}~\bibnamefont{Tolokonnikov}},
  \bibinfo{journal}{Nuclear Physics A} \textbf{\bibinfo{volume}{750}},
  \bibinfo{pages}{409} (\bibinfo{year}{2005}).

\bibitem[{\citenamefont{Grill et~al.}(2011)\citenamefont{Grill, Margueron, and
  Sandulescu}}]{gri11}
\bibinfo{author}{\bibfnamefont{F.}~\bibnamefont{Grill}},
  \bibinfo{author}{\bibfnamefont{J.}~\bibnamefont{Margueron}},
  \bibnamefont{and}
  \bibinfo{author}{\bibfnamefont{N.}~\bibnamefont{Sandulescu}},
  \bibinfo{journal}{Phys. Rev. C} \textbf{\bibinfo{volume}{84}},
  \bibinfo{pages}{065801} (\bibinfo{year}{2011}),
  \urlprefix\url{http://link.aps.org/doi/10.1103/PhysRevC.84.065801}.

\bibitem[{\citenamefont{Baldo et~al.}(2006)\citenamefont{Baldo, Saperstein, and
  Tolokonnikov}}]{bal06}
\bibinfo{author}{\bibfnamefont{M.}~\bibnamefont{Baldo}},
  \bibinfo{author}{\bibfnamefont{E.}~\bibnamefont{Saperstein}},
  \bibnamefont{and}
  \bibinfo{author}{\bibfnamefont{S.}~\bibnamefont{Tolokonnikov}},
  \bibinfo{journal}{Nuclear Physics A} \textbf{\bibinfo{volume}{775}},
  \bibinfo{pages}{235} (\bibinfo{year}{2006}).

\bibitem[{\citenamefont{Brack et~al.}(1985)\citenamefont{Brack, Guet, and
  H{\aa}kansson}}]{brack1985selfconsistent}
\bibinfo{author}{\bibfnamefont{M.}~\bibnamefont{Brack}},
  \bibinfo{author}{\bibfnamefont{C.}~\bibnamefont{Guet}}, \bibnamefont{and}
  \bibinfo{author}{\bibfnamefont{H.-B.} \bibnamefont{H{\aa}kansson}},
  \bibinfo{journal}{Physics Reports} \textbf{\bibinfo{volume}{123}},
  \bibinfo{pages}{275} (\bibinfo{year}{1985}).

\bibitem[{\citenamefont{Pearson et~al.}(2015)\citenamefont{Pearson, Chamel,
  Pastore, and Goriely}}]{pea15}
\bibinfo{author}{\bibfnamefont{J.}~\bibnamefont{Pearson}},
  \bibinfo{author}{\bibfnamefont{N.}~\bibnamefont{Chamel}},
  \bibinfo{author}{\bibfnamefont{A.}~\bibnamefont{Pastore}}, \bibnamefont{and}
  \bibinfo{author}{\bibfnamefont{S.}~\bibnamefont{Goriely}},
  \bibinfo{journal}{Physical Review C} \textbf{\bibinfo{volume}{91}},
  \bibinfo{pages}{018801} (\bibinfo{year}{2015}).

\bibitem[{\citenamefont{Chamel et~al.}(2007)\citenamefont{Chamel, Naimi, Khan,
  and Margueron}}]{cham07}
\bibinfo{author}{\bibfnamefont{N.}~\bibnamefont{Chamel}},
  \bibinfo{author}{\bibfnamefont{S.}~\bibnamefont{Naimi}},
  \bibinfo{author}{\bibfnamefont{E.}~\bibnamefont{Khan}}, \bibnamefont{and}
  \bibinfo{author}{\bibfnamefont{J.}~\bibnamefont{Margueron}},
  \bibinfo{journal}{Phys. Rev. C} \textbf{\bibinfo{volume}{75}},
  \bibinfo{pages}{055806} (\bibinfo{year}{2007}),
  \urlprefix\url{http://link.aps.org/doi/10.1103/PhysRevC.75.055806}.

\bibitem[{\citenamefont{Onsi et~al.}(2008)\citenamefont{Onsi, Dutta, Chatri,
  Goriely, Chamel, and Pearson}}]{onsi2008semi}
\bibinfo{author}{\bibfnamefont{M.}~\bibnamefont{Onsi}},
  \bibinfo{author}{\bibfnamefont{A.}~\bibnamefont{Dutta}},
  \bibinfo{author}{\bibfnamefont{H.}~\bibnamefont{Chatri}},
  \bibinfo{author}{\bibfnamefont{S.}~\bibnamefont{Goriely}},
  \bibinfo{author}{\bibfnamefont{N.}~\bibnamefont{Chamel}}, \bibnamefont{and}
  \bibinfo{author}{\bibfnamefont{J.}~\bibnamefont{Pearson}},
  \bibinfo{journal}{Physical Review C} \textbf{\bibinfo{volume}{77}},
  \bibinfo{pages}{065805} (\bibinfo{year}{2008}).

\bibitem[{\citenamefont{Pastore}(2015{\natexlab{a}})}]{pas15a}
\bibinfo{author}{\bibfnamefont{A.}~\bibnamefont{Pastore}},
  \bibinfo{journal}{Phys. Rev. C} \textbf{\bibinfo{volume}{91}},
  \bibinfo{pages}{015809} (\bibinfo{year}{2015}{\natexlab{a}}),
  \urlprefix\url{http://link.aps.org/doi/10.1103/PhysRevC.91.015809}.

\bibitem[{\citenamefont{Pearson et~al.}(2012)\citenamefont{Pearson, Chamel,
  Goriely, and Ducoin}}]{pea12}
\bibinfo{author}{\bibfnamefont{J.}~\bibnamefont{Pearson}},
  \bibinfo{author}{\bibfnamefont{N.}~\bibnamefont{Chamel}},
  \bibinfo{author}{\bibfnamefont{S.}~\bibnamefont{Goriely}}, \bibnamefont{and}
  \bibinfo{author}{\bibfnamefont{C.}~\bibnamefont{Ducoin}},
  \bibinfo{journal}{Physical Review C} \textbf{\bibinfo{volume}{85}},
  \bibinfo{pages}{065803} (\bibinfo{year}{2012}).

\bibitem[{\citenamefont{Chamel et~al.}(2015)\citenamefont{Chamel, Fantina,
  Zdunik, and Haensel}}]{chamel2015neutron}
\bibinfo{author}{\bibfnamefont{N.}~\bibnamefont{Chamel}},
  \bibinfo{author}{\bibfnamefont{A.}~\bibnamefont{Fantina}},
  \bibinfo{author}{\bibfnamefont{J.~L.} \bibnamefont{Zdunik}},
  \bibnamefont{and} \bibinfo{author}{\bibfnamefont{P.}~\bibnamefont{Haensel}},
  \bibinfo{journal}{Physical Review C} \textbf{\bibinfo{volume}{91}},
  \bibinfo{pages}{055803} (\bibinfo{year}{2015}).

\bibitem[{\citenamefont{Grill et~al.}(2012)\citenamefont{Grill,
  Provid{\^e}ncia, and Avancini}}]{grill2012neutron}
\bibinfo{author}{\bibfnamefont{F.}~\bibnamefont{Grill}},
  \bibinfo{author}{\bibfnamefont{C.}~\bibnamefont{Provid{\^e}ncia}},
  \bibnamefont{and} \bibinfo{author}{\bibfnamefont{S.~S.}
  \bibnamefont{Avancini}}, \bibinfo{journal}{Physical Review C}
  \textbf{\bibinfo{volume}{85}}, \bibinfo{pages}{055808}
  (\bibinfo{year}{2012}).

\bibitem[{\citenamefont{Grill et~al.}(2014)\citenamefont{Grill, Pais,
  Provid{\^e}ncia, Vida{\~n}a, and Avancini}}]{grill2014equation}
\bibinfo{author}{\bibfnamefont{F.}~\bibnamefont{Grill}},
  \bibinfo{author}{\bibfnamefont{H.}~\bibnamefont{Pais}},
  \bibinfo{author}{\bibfnamefont{C.}~\bibnamefont{Provid{\^e}ncia}},
  \bibinfo{author}{\bibfnamefont{I.}~\bibnamefont{Vida{\~n}a}},
  \bibnamefont{and} \bibinfo{author}{\bibfnamefont{S.~S.}
  \bibnamefont{Avancini}}, \bibinfo{journal}{Physical Review C}
  \textbf{\bibinfo{volume}{90}}, \bibinfo{pages}{045803}
  (\bibinfo{year}{2014}).

\bibitem[{\citenamefont{Margueron et~al.}(2007)\citenamefont{Margueron,
  Van~Giai, and Sandulescu}}]{mar07}
\bibinfo{author}{\bibfnamefont{J.}~\bibnamefont{Margueron}},
  \bibinfo{author}{\bibfnamefont{N.}~\bibnamefont{Van~Giai}}, \bibnamefont{and}
  \bibinfo{author}{\bibfnamefont{N.}~\bibnamefont{Sandulescu}}
  (\bibinfo{year}{2007}), \eprint{0711.0106}.

\bibitem[{\citenamefont{Chamel}(2005)}]{chamel2005band}
\bibinfo{author}{\bibfnamefont{N.}~\bibnamefont{Chamel}},
  \bibinfo{journal}{Nuclear Physics A} \textbf{\bibinfo{volume}{747}},
  \bibinfo{pages}{109} (\bibinfo{year}{2005}).

\bibitem[{\citenamefont{G{\"o}gelein and
  M{\"u}ther}(2007)}]{gogelein2007nuclear}
\bibinfo{author}{\bibfnamefont{P.}~\bibnamefont{G{\"o}gelein}}
  \bibnamefont{and}
  \bibinfo{author}{\bibfnamefont{H.}~\bibnamefont{M{\"u}ther}},
  \bibinfo{journal}{Physical Review C} \textbf{\bibinfo{volume}{76}},
  \bibinfo{pages}{024312} (\bibinfo{year}{2007}).

\bibitem[{\citenamefont{Chamel}(2012)}]{chamel2012neutron}
\bibinfo{author}{\bibfnamefont{N.}~\bibnamefont{Chamel}},
  \bibinfo{journal}{Physical Review C} \textbf{\bibinfo{volume}{85}},
  \bibinfo{pages}{035801} (\bibinfo{year}{2012}).

\bibitem[{\citenamefont{Schuetrumpf and Nazarewicz}(2015)}]{sch15}
\bibinfo{author}{\bibfnamefont{B.}~\bibnamefont{Schuetrumpf}} \bibnamefont{and}
  \bibinfo{author}{\bibfnamefont{W.}~\bibnamefont{Nazarewicz}},
  \bibinfo{journal}{Physical Review C} \textbf{\bibinfo{volume}{92}},
  \bibinfo{pages}{045806} (\bibinfo{year}{2015}).

\bibitem[{\citenamefont{Jin et~al.}(2017)\citenamefont{Jin, Bulgac, Roche, and
  Wlaz\l{}owski}}]{shi17}
\bibinfo{author}{\bibfnamefont{S.}~\bibnamefont{Jin}},
  \bibinfo{author}{\bibfnamefont{A.}~\bibnamefont{Bulgac}},
  \bibinfo{author}{\bibfnamefont{K.}~\bibnamefont{Roche}}, \bibnamefont{and}
  \bibinfo{author}{\bibfnamefont{G.}~\bibnamefont{Wlaz\l{}owski}},
  \bibinfo{journal}{Phys. Rev. C} \textbf{\bibinfo{volume}{95}},
  \bibinfo{pages}{044302} (\bibinfo{year}{2017}),
  \urlprefix\url{https://link.aps.org/doi/10.1103/PhysRevC.95.044302}.

\bibitem[{\citenamefont{Messiah}(1958)}]{messiah1958quantum}
\bibinfo{author}{\bibfnamefont{A.}~\bibnamefont{Messiah}},
  \emph{\bibinfo{title}{Quantum mechanics. vol. i..}} (\bibinfo{year}{1958}).

\bibitem[{\citenamefont{Davesne et~al.}(2016)\citenamefont{Davesne, Pastore,
  and Navarro}}]{davesne2016extended}
\bibinfo{author}{\bibfnamefont{D.}~\bibnamefont{Davesne}},
  \bibinfo{author}{\bibfnamefont{A.}~\bibnamefont{Pastore}}, \bibnamefont{and}
  \bibinfo{author}{\bibfnamefont{J.}~\bibnamefont{Navarro}},
  \bibinfo{journal}{Astronomy \& Astrophysics} \textbf{\bibinfo{volume}{585}},
  \bibinfo{pages}{A83} (\bibinfo{year}{2016}).

\bibitem[{\citenamefont{Pastore et~al.}(2011)\citenamefont{Pastore, Baroni, and
  Losa}}]{pas11}
\bibinfo{author}{\bibfnamefont{A.}~\bibnamefont{Pastore}},
  \bibinfo{author}{\bibfnamefont{S.}~\bibnamefont{Baroni}}, \bibnamefont{and}
  \bibinfo{author}{\bibfnamefont{C.}~\bibnamefont{Losa}},
  \bibinfo{journal}{Phys. Rev. C} \textbf{\bibinfo{volume}{84}},
  \bibinfo{pages}{065807} (\bibinfo{year}{2011}).

\bibitem[{\citenamefont{Bastos and O¿Hagan}(2009)}]{bastos2009diagnostics}
\bibinfo{author}{\bibfnamefont{L.~S.} \bibnamefont{Bastos}} \bibnamefont{and}
  \bibinfo{author}{\bibfnamefont{A.}~\bibnamefont{O¿Hagan}},
  \bibinfo{journal}{Technometrics} \textbf{\bibinfo{volume}{51}},
  \bibinfo{pages}{425} (\bibinfo{year}{2009}).

\bibitem[{\citenamefont{Levy and Steinberg}(2010)}]{levy2010computer}
\bibinfo{author}{\bibfnamefont{S.}~\bibnamefont{Levy}} \bibnamefont{and}
  \bibinfo{author}{\bibfnamefont{D.~M.} \bibnamefont{Steinberg}},
  \bibinfo{journal}{AStA Advances in Statistical Analysis}
  \textbf{\bibinfo{volume}{94}}, \bibinfo{pages}{311} (\bibinfo{year}{2010}).

\bibitem[{\citenamefont{Gulminelli and Raduta}(2015)}]{gulminelli2015unified}
\bibinfo{author}{\bibfnamefont{F.}~\bibnamefont{Gulminelli}} \bibnamefont{and}
  \bibinfo{author}{\bibfnamefont{A.~R.} \bibnamefont{Raduta}},
  \bibinfo{journal}{Physical Review C} \textbf{\bibinfo{volume}{92}},
  \bibinfo{pages}{055803} (\bibinfo{year}{2015}).

\bibitem[{\citenamefont{Ring and Schuck}(1980)}]{Book:Ring1980}
\bibinfo{author}{\bibfnamefont{P.}~\bibnamefont{Ring}} \bibnamefont{and}
  \bibinfo{author}{\bibfnamefont{P.}~\bibnamefont{Schuck}},
  \emph{\bibinfo{title}{{The Nuclear Many-Body Problem}}}
  (\bibinfo{publisher}{Springer-Verlag}, \bibinfo{year}{1980}).

\bibitem[{\citenamefont{Pastore}(2012)}]{pas12}
\bibinfo{author}{\bibfnamefont{A.}~\bibnamefont{Pastore}},
  \bibinfo{journal}{Phys. Rev. C} \textbf{\bibinfo{volume}{86}},
  \bibinfo{pages}{065802} (\bibinfo{year}{2012}).

\bibitem[{\citenamefont{Pastore et~al.}(2013)\citenamefont{Pastore, Margueron,
  Schuck, and Vi{\~n}as}}]{pastore2013pairing}
\bibinfo{author}{\bibfnamefont{A.}~\bibnamefont{Pastore}},
  \bibinfo{author}{\bibfnamefont{J.}~\bibnamefont{Margueron}},
  \bibinfo{author}{\bibfnamefont{P.}~\bibnamefont{Schuck}}, \bibnamefont{and}
  \bibinfo{author}{\bibfnamefont{X.}~\bibnamefont{Vi{\~n}as}},
  \bibinfo{journal}{Physical Review C} \textbf{\bibinfo{volume}{88}},
  \bibinfo{pages}{034314} (\bibinfo{year}{2013}).

\bibitem[{\citenamefont{Pastore}(2015{\natexlab{b}})}]{pastore2015pairing}
\bibinfo{author}{\bibfnamefont{A.}~\bibnamefont{Pastore}},
  \bibinfo{journal}{Physical Review C} \textbf{\bibinfo{volume}{91}},
  \bibinfo{pages}{015809} (\bibinfo{year}{2015}{\natexlab{b}}).

\bibitem[{\citenamefont{Chabanat et~al.}(1997)\citenamefont{Chabanat, Bonche,
  Haensel, Meyer, and Schaeffer}}]{cha97}
\bibinfo{author}{\bibfnamefont{E.}~\bibnamefont{Chabanat}},
  \bibinfo{author}{\bibfnamefont{P.}~\bibnamefont{Bonche}},
  \bibinfo{author}{\bibfnamefont{P.}~\bibnamefont{Haensel}},
  \bibinfo{author}{\bibfnamefont{J.}~\bibnamefont{Meyer}}, \bibnamefont{and}
  \bibinfo{author}{\bibfnamefont{R.}~\bibnamefont{Schaeffer}},
  \bibinfo{journal}{Nuclear Physics A} \textbf{\bibinfo{volume}{627}},
  \bibinfo{pages}{710} (\bibinfo{year}{1997}).

\bibitem[{\citenamefont{Bennaceur and Dobaczewski}(2005)}]{ben05}
\bibinfo{author}{\bibfnamefont{K.}~\bibnamefont{Bennaceur}} \bibnamefont{and}
  \bibinfo{author}{\bibfnamefont{J.}~\bibnamefont{Dobaczewski}},
  \bibinfo{journal}{Computer physics communications}
  \textbf{\bibinfo{volume}{168}}, \bibinfo{pages}{96} (\bibinfo{year}{2005}).

\bibitem[{\citenamefont{Garrido et~al.}(1999)\citenamefont{Garrido, Sarriguren,
  De~Guerra, and Schuck}}]{gar99}
\bibinfo{author}{\bibfnamefont{E.}~\bibnamefont{Garrido}},
  \bibinfo{author}{\bibfnamefont{P.}~\bibnamefont{Sarriguren}},
  \bibinfo{author}{\bibfnamefont{E.~M.} \bibnamefont{De~Guerra}},
  \bibnamefont{and} \bibinfo{author}{\bibfnamefont{P.}~\bibnamefont{Schuck}},
  \bibinfo{journal}{Physical Review C} \textbf{\bibinfo{volume}{60}},
  \bibinfo{pages}{064312} (\bibinfo{year}{1999}).

\bibitem[{\citenamefont{Baroni et~al.}(2010)\citenamefont{Baroni, Macchiavelli,
  and Schwenk}}]{baroni2010partial}
\bibinfo{author}{\bibfnamefont{S.}~\bibnamefont{Baroni}},
  \bibinfo{author}{\bibfnamefont{A.~O.} \bibnamefont{Macchiavelli}},
  \bibnamefont{and} \bibinfo{author}{\bibfnamefont{A.}~\bibnamefont{Schwenk}},
  \bibinfo{journal}{Physical Review C} \textbf{\bibinfo{volume}{81}},
  \bibinfo{pages}{064308} (\bibinfo{year}{2010}).

\bibitem[{\citenamefont{Bulgac and Yu}(2002)}]{bul02}
\bibinfo{author}{\bibfnamefont{A.}~\bibnamefont{Bulgac}} \bibnamefont{and}
  \bibinfo{author}{\bibfnamefont{Y.}~\bibnamefont{Yu}},
  \bibinfo{journal}{Physical review letters} \textbf{\bibinfo{volume}{88}},
  \bibinfo{pages}{042504} (\bibinfo{year}{2002}).

\bibitem[{\citenamefont{Gandolfi et~al.}(2008)\citenamefont{Gandolfi,
  Illarionov, Fantoni, Pederiva, and Schmidt}}]{gandolfi2008equation}
\bibinfo{author}{\bibfnamefont{S.}~\bibnamefont{Gandolfi}},
  \bibinfo{author}{\bibfnamefont{A.~Y.} \bibnamefont{Illarionov}},
  \bibinfo{author}{\bibfnamefont{S.}~\bibnamefont{Fantoni}},
  \bibinfo{author}{\bibfnamefont{F.}~\bibnamefont{Pederiva}}, \bibnamefont{and}
  \bibinfo{author}{\bibfnamefont{K.~E.} \bibnamefont{Schmidt}},
  \bibinfo{journal}{Physical review letters} \textbf{\bibinfo{volume}{101}},
  \bibinfo{pages}{132501} (\bibinfo{year}{2008}).

\bibitem[{\citenamefont{Goriely et~al.}(2016)\citenamefont{Goriely, Chamel, and
  Pearson}}]{goriely2016further}
\bibinfo{author}{\bibfnamefont{S.}~\bibnamefont{Goriely}},
  \bibinfo{author}{\bibfnamefont{N.}~\bibnamefont{Chamel}}, \bibnamefont{and}
  \bibinfo{author}{\bibfnamefont{J.}~\bibnamefont{Pearson}},
  \bibinfo{journal}{Physical Review C} \textbf{\bibinfo{volume}{93}},
  \bibinfo{pages}{034337} (\bibinfo{year}{2016}).

\bibitem[{\citenamefont{Shapiro and Teukolsky}(1983)}]{sha93}
\bibinfo{author}{\bibfnamefont{S.~L.} \bibnamefont{Shapiro}} \bibnamefont{and}
  \bibinfo{author}{\bibfnamefont{S.~A.} \bibnamefont{Teukolsky}},
  \bibinfo{journal}{Black Holes, White Dwarfs, and Neutron Stars: The Physics
  of Compact Objects} pp. \bibinfo{pages}{335--369} (\bibinfo{year}{1983}).

\bibitem[{\citenamefont{Salpeter}(1961)}]{sal61}
\bibinfo{author}{\bibfnamefont{E.~E.} \bibnamefont{Salpeter}},
  \bibinfo{journal}{The Astrophysical Journal} \textbf{\bibinfo{volume}{134}},
  \bibinfo{pages}{669} (\bibinfo{year}{1961}).

\bibitem[{\citenamefont{Maruyama et~al.}(2005)\citenamefont{Maruyama, Tatsumi,
  Voskresensky, Tanigawa, and Chiba}}]{maruyama2005nuclear}
\bibinfo{author}{\bibfnamefont{T.}~\bibnamefont{Maruyama}},
  \bibinfo{author}{\bibfnamefont{T.}~\bibnamefont{Tatsumi}},
  \bibinfo{author}{\bibfnamefont{D.~N.} \bibnamefont{Voskresensky}},
  \bibinfo{author}{\bibfnamefont{T.}~\bibnamefont{Tanigawa}}, \bibnamefont{and}
  \bibinfo{author}{\bibfnamefont{S.}~\bibnamefont{Chiba}},
  \bibinfo{journal}{Physical Review C} \textbf{\bibinfo{volume}{72}},
  \bibinfo{pages}{015802} (\bibinfo{year}{2005}).

\bibitem[{\citenamefont{Watanabe and Iida}(2003)}]{watanabe2003electron}
\bibinfo{author}{\bibfnamefont{G.}~\bibnamefont{Watanabe}} \bibnamefont{and}
  \bibinfo{author}{\bibfnamefont{K.}~\bibnamefont{Iida}},
  \bibinfo{journal}{Physical Review C} \textbf{\bibinfo{volume}{68}},
  \bibinfo{pages}{045801} (\bibinfo{year}{2003}).

\bibitem[{\citenamefont{Alcain et~al.}(2014)\citenamefont{Alcain, Molinelli,
  Nichols, and Dorso}}]{alcain2014effect}
\bibinfo{author}{\bibfnamefont{P.}~\bibnamefont{Alcain}},
  \bibinfo{author}{\bibfnamefont{P.~G.} \bibnamefont{Molinelli}},
  \bibinfo{author}{\bibfnamefont{J.}~\bibnamefont{Nichols}}, \bibnamefont{and}
  \bibinfo{author}{\bibfnamefont{C.}~\bibnamefont{Dorso}},
  \bibinfo{journal}{Physical Review C} \textbf{\bibinfo{volume}{89}},
  \bibinfo{pages}{055801} (\bibinfo{year}{2014}).

\bibitem[{\citenamefont{Bonche and Vautherin}(1981)}]{bon81}
\bibinfo{author}{\bibfnamefont{P.}~\bibnamefont{Bonche}} \bibnamefont{and}
  \bibinfo{author}{\bibfnamefont{D.}~\bibnamefont{Vautherin}},
  \bibinfo{journal}{Nuclear Physics A} \textbf{\bibinfo{volume}{372}},
  \bibinfo{pages}{496} (\bibinfo{year}{1981}).

\bibitem[{\citenamefont{Abramowitz and Stegun}(1964)}]{abramowitz1964handbook}
\bibinfo{author}{\bibfnamefont{M.}~\bibnamefont{Abramowitz}} \bibnamefont{and}
  \bibinfo{author}{\bibfnamefont{I.~A.} \bibnamefont{Stegun}},
  \emph{\bibinfo{title}{Handbook of mathematical functions: with formulas,
  graphs, and mathematical tables}}, vol.~\bibinfo{volume}{55}
  (\bibinfo{publisher}{Courier Corporation}, \bibinfo{year}{1964}).

\bibitem[{\citenamefont{Strutinsky}(1967)}]{strutinsky1967shell}
\bibinfo{author}{\bibfnamefont{V.}~\bibnamefont{Strutinsky}},
  \bibinfo{journal}{Nuclear Physics A} \textbf{\bibinfo{volume}{95}},
  \bibinfo{pages}{420} (\bibinfo{year}{1967}).

\bibitem[{\citenamefont{Pastore et~al.}(2016)\citenamefont{Pastore, Shelley,
  and Diget}}]{pastore2016pos}
\bibinfo{author}{\bibfnamefont{A.}~\bibnamefont{Pastore}},
  \bibinfo{author}{\bibfnamefont{M.}~\bibnamefont{Shelley}}, \bibnamefont{and}
  \bibinfo{author}{\bibfnamefont{C.~A.} \bibnamefont{Diget}},
  \bibinfo{journal}{PoS (INPC2016)} p. \bibinfo{pages}{145}
  (\bibinfo{year}{2016}).

\bibitem[{\citenamefont{Aymard et~al.}(2014)\citenamefont{Aymard, Gulminelli,
  and Margueron}}]{aymard2014medium}
\bibinfo{author}{\bibfnamefont{F.}~\bibnamefont{Aymard}},
  \bibinfo{author}{\bibfnamefont{F.}~\bibnamefont{Gulminelli}},
  \bibnamefont{and}
  \bibinfo{author}{\bibfnamefont{J.}~\bibnamefont{Margueron}},
  \bibinfo{journal}{Physical Review C} \textbf{\bibinfo{volume}{89}},
  \bibinfo{pages}{065807} (\bibinfo{year}{2014}).

\bibitem[{\citenamefont{Toivanen et~al.}(2008)\citenamefont{Toivanen,
  Dobaczewski, Kortelainen, and Mizuyama}}]{toivanen2008error}
\bibinfo{author}{\bibfnamefont{J.}~\bibnamefont{Toivanen}},
  \bibinfo{author}{\bibfnamefont{J.}~\bibnamefont{Dobaczewski}},
  \bibinfo{author}{\bibfnamefont{M.}~\bibnamefont{Kortelainen}},
  \bibnamefont{and} \bibinfo{author}{\bibfnamefont{K.}~\bibnamefont{Mizuyama}},
  \bibinfo{journal}{Physical Review C} \textbf{\bibinfo{volume}{78}},
  \bibinfo{pages}{034306} (\bibinfo{year}{2008}).

\bibitem[{\citenamefont{Gao et~al.}(2013)\citenamefont{Gao, Dobaczewski,
  Kortelainen, Toivanen, Tarpanov et~al.}}]{gao2013propagation}
\bibinfo{author}{\bibfnamefont{Y.}~\bibnamefont{Gao}},
  \bibinfo{author}{\bibfnamefont{J.}~\bibnamefont{Dobaczewski}},
  \bibinfo{author}{\bibfnamefont{M.}~\bibnamefont{Kortelainen}},
  \bibinfo{author}{\bibfnamefont{J.}~\bibnamefont{Toivanen}},
  \bibinfo{author}{\bibfnamefont{D.}~\bibnamefont{Tarpanov}},
  \bibnamefont{et~al.}, \bibinfo{journal}{Physical Review C}
  \textbf{\bibinfo{volume}{87}}, \bibinfo{pages}{034324}
  (\bibinfo{year}{2013}).

\bibitem[{\citenamefont{Haverinen and
  Kortelainen}(2016)}]{haverinen2016uncertainty}
\bibinfo{author}{\bibfnamefont{T.}~\bibnamefont{Haverinen}} \bibnamefont{and}
  \bibinfo{author}{\bibfnamefont{M.}~\bibnamefont{Kortelainen}},
  \bibinfo{journal}{arXiv preprint arXiv:1611.00597}  (\bibinfo{year}{2016}).

\bibitem[{\citenamefont{Goriely et~al.}(2010)\citenamefont{Goriely, Chamel, and
  Pearson}}]{Gor10}
\bibinfo{author}{\bibfnamefont{S.}~\bibnamefont{Goriely}},
  \bibinfo{author}{\bibfnamefont{N.}~\bibnamefont{Chamel}}, \bibnamefont{and}
  \bibinfo{author}{\bibfnamefont{J.~M.} \bibnamefont{Pearson}},
  \bibinfo{journal}{Phys. Rev. C} \textbf{\bibinfo{volume}{82}},
  \bibinfo{pages}{035804} (\bibinfo{year}{2010}),
  \urlprefix\url{http://link.aps.org/doi/10.1103/PhysRevC.82.035804}.

\bibitem[{\citenamefont{Utama et~al.}(2016)\citenamefont{Utama, Piekarewicz,
  and Prosper}}]{utama2016nuclear}
\bibinfo{author}{\bibfnamefont{R.}~\bibnamefont{Utama}},
  \bibinfo{author}{\bibfnamefont{J.}~\bibnamefont{Piekarewicz}},
  \bibnamefont{and} \bibinfo{author}{\bibfnamefont{H.}~\bibnamefont{Prosper}},
  \bibinfo{journal}{Physical Review C} \textbf{\bibinfo{volume}{93}},
  \bibinfo{pages}{014311} (\bibinfo{year}{2016}).

\bibitem[{\citenamefont{Sharma et~al.}(2015)\citenamefont{Sharma, Centelles,
  Vinas, Baldo, and Burgio}}]{sha15}
\bibinfo{author}{\bibfnamefont{B.}~\bibnamefont{Sharma}},
  \bibinfo{author}{\bibfnamefont{M.}~\bibnamefont{Centelles}},
  \bibinfo{author}{\bibfnamefont{X.}~\bibnamefont{Vinas}},
  \bibinfo{author}{\bibfnamefont{M.}~\bibnamefont{Baldo}}, \bibnamefont{and}
  \bibinfo{author}{\bibfnamefont{G.}~\bibnamefont{Burgio}},
  \bibinfo{journal}{Astronomy \& Astrophysics} \textbf{\bibinfo{volume}{584}},
  \bibinfo{pages}{A103} (\bibinfo{year}{2015}).

\bibitem[{\citenamefont{Roca-Maza and Piekarewicz}(2008)}]{roca2008impact}
\bibinfo{author}{\bibfnamefont{X.}~\bibnamefont{Roca-Maza}} \bibnamefont{and}
  \bibinfo{author}{\bibfnamefont{J.}~\bibnamefont{Piekarewicz}},
  \bibinfo{journal}{Physical Review C} \textbf{\bibinfo{volume}{78}},
  \bibinfo{pages}{025807} (\bibinfo{year}{2008}).

\bibitem[{\citenamefont{Pearson et~al.}(2011)\citenamefont{Pearson, Goriely,
  and Chamel}}]{pearson2011properties}
\bibinfo{author}{\bibfnamefont{J.}~\bibnamefont{Pearson}},
  \bibinfo{author}{\bibfnamefont{S.}~\bibnamefont{Goriely}}, \bibnamefont{and}
  \bibinfo{author}{\bibfnamefont{N.}~\bibnamefont{Chamel}},
  \bibinfo{journal}{Physical Review C} \textbf{\bibinfo{volume}{83}},
  \bibinfo{pages}{065810} (\bibinfo{year}{2011}).

\end{thebibliography}

\end{document}